\documentclass[aps,11pt,preprintnumbers,nofootinbib,floatfix]{revtex4}

\usepackage{graphicx}
\usepackage{epsfig}
\usepackage{dcolumn}
\usepackage{subfigure}
\usepackage{multirow}
\usepackage{amsmath}
\usepackage{amssymb}

\begin{document}

\title{Gauss-Bonnet holographic superconductors in exponential nonlinear electrodynamics}


\author{Cao H. Nam}
\email{hncao@yonsei.ac.kr} \affiliation{
Institute of Research and Development, Duy Tan
University, Da Nang 550000, Vietnam}
\date{\today}

\begin{abstract}%

The low-energy limits of the string theory lead to the higher-order curvature corrections for Einstein gravity. Also, they give the higher-order derivative corrections for the Maxwell or linear electrodynamics, which suggests the nonlinear electrodynamics. Inspired by this, in this paper we investigate $d$-dimensional holographic superconductors in the probe limit in the framework of Einstein-Gauss-Bonnet gravity and exponential nonlinear electrodynamics. Based on the Sturm-Liouville eigenvalue method, we compute the critical temperature, the condensation value, and the critical exponent. It is observed that the critical temperature decreases when the Gauss-Bonnet (GB) parameter or the nonlinear parameter increases, but it increases with the higher dimension of the spacetime at the efficiently low charge density. In addition, we found that the condensation value becomes larger as increasing the GB parameter, the nonlinear parameter as well as the spacetime dimension. Finally, we calculate the optical conductivity and study the effects of the GB term and exponential nonlinear electrodynamics on superconducting energy gap.
\end{abstract}

\maketitle

\section{Introduction}

The BCS theory provides a microscopic description with great accuracy for conventional or low temperature superconductors \cite{Schrieffer1957}. Here, the occurrence of the superconducting state is a result of the spontaneously broken $\mathrm{U}(1)$ symmetry due to the condensation of Cooper pairs which are formed by the attractive interaction coming from the virtual exchange of the phonons. However, the BCS theory fails in explaining of the pairing mechanism in the high temperature superconductors in which the system is strongly coupled. Since, understanding the mechanism of the high temperature superconductors has been one of the most important problems in condensed-matter physics and it requires new theoretical approaches.

The AdS/CFT correspondence, proposed firstly by Maldacena, indicates a correspondence between a weakly coupled garvity theory in a $d$-dimensional anti-de Sitter (AdS) spacetime and a strongly coupled conformal field theory (CFT) sitting on the $(d-1)$-dimesnional boundary of the bulk AdS spacetime \cite{Maldacena,Witten,Gubser,Aharony}. Thus, the AdS/CFT correspondence provides a powerful tool to calculate the properties of the strongly coupled systems by exploiting the weakly coupled gravitational systems. By the computational power of the AdS/CFT correspondence, it has been widely used to study open problems in condensed-matter physics, especially the high temperature superconductors. 

It was proposed by Hartnoll et al. that $s$-wave superconductors can be described by the gravitational system in one higher dimension in the probe limit, well-known as holographic superconductors \cite{Horowitz2008a,Horowitz2008b}. Basically, the holographic model of $s$-wave superconductors contains a planar AdS-Schwarzschild black hole and a charged scalar field coupled minimally to a Maxwell field. Here, the Maxwell field and charged scalar field describe the $\mathrm{U}(1)$ symmetry and the charged scalar operator in the dual field theory, respectively. It was found that below a critical temperature there has the formation of a scalar hair, indicating a condensation in the dual field theory, due to the breaking of the $\mathrm{U}(1)$ symmetry near the event horizon \cite{Gubser2005,Gubser2008}. This corresponds to a phase transition from a black hole with no hair (normal/conducting phase in the dual field theory) to a black hole with scalar hair (superconducting phase in the dual field theory). Following the remarkable work of Hartnoll et al., holographic superconductors have attracted a lot of attention and have been investigated extensively in the literature \cite{Roberts2008,Rodriguez-Gomez2010,Therrien2010,Yang2010,Wu2011,Zong2011,Zhang2011,Yang2015}. Also, other works have been devoted to study $s$-wave holographic superconductors concerning the modifications of the two gauge fields \cite{Maggiore2014,Rogatko2014,Rogatko2015a,Rogatko2015b,Pan-Liu2017}.

All of the above mentioned works of the holographic superconductors are considered in the framework of the linear or Maxwell electrodynamics. Interestingly, the low-energy limit of the heterotic string theory
\cite{Natsuume1994,Padi2007,Sun2008,Szepietowski2009,Pastras2009} and calculating one-loop approximation of QED \cite{Delbourgo1996} lead naturally to the nonlinear electrodynamics. Thus, the nonlinear electrodynamics has drawn attention in investigating holographic superconductors \cite{Pan2011,Gangopadhyay2012,Roychowdhury2012,Jing2013,Lala2013,Lai2015,
Sheykhi2016,Ghorai2016,Jiang2016,Salahi2016,Liu2016,Ghazanfari2018,Asl2018}.

Exponential nonlinear electrodynamics is one of Born-Infeld-type electrodynamics forms, besides Born-Infeld and logarithmic nonlinear electrodynamics. It should be emphasized here that the forms of Born-Infeld-type electrodynamics possess the special properties such as the absence of the shock waves, the birefringence phenomena \cite{Boillat1970} as well as enjoying an electric-magnetic duality \cite{Gibbons1995}. Another property of the exponential nonlinear electrodynamics is that it would reduce to Maxwell electrodynamics in the region of the weak field or the small nonlinear parameter. The exponential form of the nonlinear electrodynamics was introduced in Ref. \cite{Hendi2012} to obtain the black hole solutions whose asymptotic properties are the same as the charged BTZ solution. It was indicated that the exponential nonlinear electrodynamics can not remove the divergency of the electric field at $r=0$, but its singularity is much weaker than the Einstein-Maxwell theory \cite{SHHendi2013,SHHendi2014}. Various charged black hole solutions have been found with the source of the exponential nonlinear electrodynamics \cite{Kazemi2014,Matsuno2015,Kruglov2016,Kruglov2017,Hajkhalili2018}. In addition, holographic superconductors with the exponential nonlinear electrodynamics have been investigated in the $(3+1)$-dimensions in the probe limit \cite{Jing2013}. Ref. \cite{Jing2013} indicated that the exponential nonlinear electrodynamics has stronger effect on the condensation of the $(3+1)$-dimensional superconductor compared to other nonlinear electrodynamics models.

It is widely believed that Einstein gravity or General Relativity (GR) is no longer
true in the regime of the very high energy or very short distance at which the corrections of
a more fundamental theory of the gravitation, e.g. quantum gravity, must be taken into account.
Einstein-Gauss-Bonnet gravity is one of the natural generalizations of GR with including the higher-order curvature corrections, because it arises from the low-energy limit of heterotic string theory \cite{Zwiebach1985,Witten1986,Gross1987,Tseytlin1987,Bento1996}. The properties of GB holographic superconductors have been investigated in many works with Maxwell electrodynamics as well as various forms of the nonlinear electrodynamics  \cite{Gregory2009,Pavan2010,Wang2010,Barclay2010,Nie2010,Cai2011,Kanno2010,Kanno2011,Pan-Chen2011,Nie2011,Barclay2011,Wang2011,Jing-Chen2012,Cai-Zhang2013,Cui-Xue2013,Yao-Jing2013,Li2014,Dev-Lala2014,Gangopadhyay2019,CHNam2019}. GB holographic superconductors with the exponential nonlinear electrodynamics for the case of the five-dimensional spacetime already has been done using matching method \cite{Dev-Lala2014}. It was found that increasing the GB coupling parameter should make the condensation harder. However, the critical exponent of the system is not affected by the GB corrections, which is consistent with mean field theory.

So far, studying GB holographic superconductors with the exponential nonlinear electrodynamics in the arbitrary spacetime dimension as well as calculating the optical conductivity have not been done yet. 
Thus, in this paper we investigate holographic superconductors in the arbitrary spacetime dimension in the context of Einstein-Gauss-Bonnet gravity and the exponential nonlinear electrodynamics. We study how the GB corrections, the nonlinear parameter as well as the spacetime dimension affect the critical temperature, condensation value, critical exponent, the optical conductivity, and superconducting energy gap. For the analytical study, we use the Sturm-Liouville eigenvalue method and then compare the analytical results to the numerical results.

This paper is organized as follows. In Sec. \ref{HDM}, we build a holographic dual model which consists of a $d$-dimensional planar GB AdS black hole background and a charged scalar field coupled minimally to a nonlinear gauge field whose Langrangian is the exponential form. In Sec. \ref{CTVCD}, we obtain the relation between the critical temperature and the charge density based on the analytical method, and then compare the analytical results with the numerical results. In Sec. \ref{CVCE}, we compute the condensation value and critical exponent of the superconducting system. In Sec. \ref{HC}, we study the behavior of the optical conductivity (as a function of the frequency) and superconducting energy gap, by turning on the fluctuations of the vector gauge field in the bulk. Finally, we make conclusions in the last section, Sec. \ref{conclu}.

\section{\label{HDM} Holographic dual model}
In this section, we will build a holographic dual model which is used for the subsequent computation of critical phenomena and conductivity. We start with a $d$-dimensional action which describes Einstein-Gauss-Bonnet gravity, a nonlinear electromagnetic field and a charged scalar field in the AdS spacetime background as follows
\begin{equation}
S=\frac{1}{16\pi}\int
d^dx\sqrt{-g}\left[R-2\Lambda+\alpha\mathcal{L}_{GB}+\mathcal{L}_{\text{m}}\right],\label{EGB-nlED-adS}
\end{equation}
where $R$ is the scalar curvature of the spacetime, $\Lambda$ is the negative cosmological constant related to the curvature radius $l$ of the AdS spacetime background as
\begin{equation}
\Lambda=-\frac{(d-1)(d-2)}{2l^2},
\end{equation}
$\mathcal{L}_{GB}$ is the GB term given as
\begin{equation}
  \mathcal{L}_{GB}=R^2-4R_{\mu\nu}R^{\mu\nu}+R_{\mu\nu\rho\lambda}R^{\mu\nu\rho\lambda},
\end{equation}
and $\alpha$ is the GB coupling parameter\footnote{The GB term appears in the
heterotic string theory at which $\alpha$ is regarded as the inverse string tension \cite{Gross1987,Bento1996}. Thus, in this paper only the case $\alpha\geq0$ is considered.}. The matter term is given as
\begin{equation}
\mathcal{L}_{\text{m}}=\frac{1}{b}\left(e^{-bF}-1\right)-|(\nabla_\mu-iqA_\mu)\psi|^2-m^2|\psi|^2,
\end{equation}
where $F\equiv\frac{F_{\mu\nu}F^{\mu\nu}}{4}$ with $F_{\mu\nu}=\partial_\mu
A_\nu-\partial_\nu A_\mu$ to be the strength tensor of the nonlinear electromagnetic field $A_\mu$, $b$ is the nonlinear parameter, and $\psi$ refers to the scalar field of the charge $q$ and the mass $m$.

Varying the action (\ref{EGB-nlED-adS}) with respect to the metric $g_{\mu\nu}$, one can find the Einstein-Gauss-Bonnet field equations as
\begin{eqnarray}
{G^\mu}_\nu+\alpha{H^\mu}_\nu-\frac{(d-1)(d-2)}{2l^2}{\delta^\mu}_\nu&=&{T^\mu}_\nu(\text{matter}),\label{Eeq}
\end{eqnarray}
where
\begin{equation}
  {H^\mu}_\nu=2\left(R{R^\mu}_\nu-2R^{\mu\sigma}R_{\sigma\nu}-2R^{\sigma\rho}{R^\mu}_{\sigma\nu\rho}+{R^\mu}_{\rho\sigma\lambda}{R_\nu}^{\rho\sigma\lambda}\right)-\frac{1}{2}{\delta^\mu}_\nu\mathcal{L}_{GB},
\end{equation}
and ${T^\mu}_\nu(\text{matter})$ is the energy-momentum tensor of the matter field. In this work, we study the system in the probe limit which means that the scalar and Maxwell fields decouple from the gravity. Since their backreaction on the spacetime geometry is ignored or ${T^\mu}_\nu(\text{matter})\simeq0$. On the other hand, in the probe limit the scalar and Maxwell fields are considered in a given spacetime background which is a planar-symmetric GB AdS black hole determined by the following metric \cite{Cai2002}
\begin{equation}
ds^2=-f(r)dt^2+\frac{dr^2}{f(r)}+r^2h_{ij}dx^idx^j,
\end{equation}
where the function $f(r)$ is given by
\begin{equation}
f(r)=\frac{r^2}{2\widetilde{\alpha}}\left[1-\sqrt{1-\frac{4\widetilde{\alpha}}{l^2}\left(1-\frac{r^{d-1}_+}{r^{d-1}}\right)}\right],
\end{equation}
with $r_+$ to be the event horizon radius, $\widetilde{\alpha}=\alpha(d-3)(d-4)$, and $h_{ij}dx^idx^j=dx^2_1+dx^2_2+...+dx^2_{d-2}$ is the line element of the $(d-2)$-dimensional planar hypersurface. Near the asymptotic region ($r\rightarrow\infty$), the function $f(r)$ becomes
\begin{equation}
f(r)=\frac{r^2}{l^2_{\text{eff}}},
\end{equation}
where
\begin{equation}
l^2_{\text{eff}}=\frac{2\widetilde{\alpha}}{1-\sqrt{1-\frac{4\widetilde{\alpha}}{l^2}}},
\end{equation}
is defined as the effective AdS radius. It should be noted here that $\widetilde{\alpha}\leq l^2/4$ must be satisfied for the theory defined. The Hawking temperature of the black hole, which is interpreted as the temperature of the dual field theory, is given by
\begin{equation}
T=\frac{f'(r_+)}{4\pi}=\frac{(d-1)r_+}{4\pi l^2}.\label{Haw-temp}
\end{equation}

By varying the above action with respect to the electromagnetic field $A_\mu$ and the scalar field $\psi$, we obtain the equations of motion as
\begin{eqnarray}
\nabla^\nu\left(e^{-bF}F_{\mu\nu}\right)+iq\left[\psi^*(\nabla_\mu-iqA_\mu)\psi-\psi(\nabla_\mu+iqA_\mu)\psi^*\right]&=&0,\nonumber\\
\left(\nabla_\mu-iqA_\mu\right)\left(\nabla^\mu-iqA^\mu\right)\psi-m^2\psi &=&0.\label{VSeq}
\end{eqnarray}
Now we find a solution of these equations with the following ansatz
\begin{equation}
A_\mu=\phi(r)\delta^t_\mu,\ \ \ \ \psi=\psi(r).
\end{equation}
By substituting this ansatz into Eq. (\ref{VSeq}), we obtain the equation for the functions $\phi(r)$ and $\psi(r)$ as
\begin{eqnarray}
 \left[1+b\phi'^2(r)\right]\phi''(r)+\frac{d-2}{r}\phi'(r)-\frac{2q^2\phi(r)}{f(r)}\psi^2(r)e^{-\frac{b}{2}\phi'^2(r)}&=&0,\label{r-phi-Eq}\\
 \psi''(r)+\left[\frac{f'(r)}{f(r)}+\frac{d-2}{r}\right]\psi'(r)+\left[\frac{q^2\phi^2(r)}{f^2(r)}-\frac{m^2}{f(r)}\right]\psi(r)&=&0,\label{r-psi-Eq}
\end{eqnarray}
where the prime is denoted the derivative with respect to the coordinate $r$. One can see that these nonlinear equations possess several scaling symmetries as
\begin{enumerate}
\item $\phi\rightarrow\lambda_1\phi,\ \ \psi\rightarrow\lambda_1\psi,\  \ q\rightarrow\lambda^{-1}_1q,\ \ b\rightarrow\lambda^{-2}_1b$.
\item $\phi\rightarrow\lambda_2\phi,\ \ \psi\rightarrow\lambda^{1/2}_2\psi,\ \ b\rightarrow\lambda^{-2}_2b\ \ \widetilde{\alpha}\rightarrow\lambda^{-1}_2\widetilde{\alpha},\ \ l\rightarrow\lambda^{-1/2}_2l,\ \ m\rightarrow\lambda^{1/2}_2m$.
\end{enumerate}
Using these scaling symmetries, one can set $q=l=1$. In order to solve Eqs. (\ref{r-phi-Eq}) and (\ref{r-psi-Eq}) and obtain the behavior of the gauge field and scalar field, first we need to impose the appropriate boundary conditions for $\phi(r)$ and $\psi(r)$ at the event horizon $r_+$. The matter fields are regular at the event horizon $r_+$, which means that their value must be finite at $r_+$. Since, the regularity condition for the matter fields leads to the following boundary conditions
\begin{equation}
\phi(r_+)=0,\  \ \psi(r_+)=\frac{f'(r_+)\psi'(r_+)}{m^2}.
\end{equation}
Solving Eqs. (\ref{r-phi-Eq}) and (\ref{r-psi-Eq}) in the the asymptotic limit ($r\rightarrow\infty$), we obtain the behavior of $\phi(r)$ and $\psi(r)$ near the AdS boundary as
\begin{eqnarray}
\phi(r)&=&\mu-\frac{\rho}{r^{d-3}},\label{phi-asy-beh}\\
\psi(r)&=&\frac{\langle\mathcal{O}_-\rangle}{r^{\Delta_-}}+\frac{\langle\mathcal{O}_+\rangle}{r^{\Delta_+}}.
\end{eqnarray}
Here, the parameters $\mu$ and $\rho$ are regarded as the chemical potential and the charge density, respectively. According to the AdS/CFT dictionary, the coefficients $\langle\mathcal{O}_-\rangle$ and $\langle\mathcal{O}_+\rangle$ are interpreted as the source and the expectation value of the condensation operator (dual to the bulk scalar field) in the dual field theory, respectively. Because the $\mathrm{U}(1)$ symmetry is required to be broken spontaneously, we should turn off the source, $\langle\mathcal{O}_-\rangle=0$, which keeps up the stability of the AdS space \cite{Roberts2008}. And, the conformal dimension $\Delta_\pm$ of $\langle\mathcal{O}_\pm\rangle$ is given as
\begin{equation}
\Delta_\pm=\frac{1}{2}\left[(d-1)\pm\sqrt{(d-1)^2+4m^2l^2_{\text{eff}}}\right],
\end{equation}
which implies the Breitenlohner-Freedman (BF) bound \cite{Freedman1982,Breitenlohner1982} for the mass of the scalar field
\begin{equation}
m^2\geq-\frac{(d-1)^2}{4l^2_{\text{eff}}}.
\end{equation}

In preparation for proceeding, let us rewrite Eqs. (\ref{r-phi-Eq}) and (\ref{r-psi-Eq}) in the new coordinate $z=\frac{r_+}{r}$ as
\begin{eqnarray}
 \left[1+\frac{bz^4}{r^2_+}\phi'^2(z)\right]\phi''(z)+\frac{2bz^3}{r^2_+}\phi'^3(z)+\frac{4-d}{z}\phi'(z)-\frac{2 r^2_+\phi(z)}{z^4f(z)}\psi^2(z)e^{-\frac{bz^4}{2r^2_+}\phi'^2(z)}&=&0,\label{z-phi-Eq}\\
 \psi''(z)+\left(\frac{f'(z)}{f(z)}+\frac{4-d}{z}\right)\psi'(z)+\frac{r^2_+}{z^4}\left(\frac{\phi^2(z)}{f^2(z)}-\frac{m^2}{f(z)}\right)\psi(z)&=&0,\label{z-psi-Eq}
\end{eqnarray}
where the prime is denoted the derivative with respect to the coordinate $z$, and
\begin{equation}
f(z)=\frac{r^2_+}{2\widetilde{\alpha}z^2}\left(1-\sqrt{1-4\widetilde{\alpha}\left(1-z^{d-1}\right)}\right).
\end{equation}

\section{\label{CTVCD} Critical temperature versus charge density}
In this section, we will find the relation between the critical temperature $T_c$ and the charge density $\rho$ for GB holographic superconductors in the exponential nonlinear electrodynamics. Analytic study for this is based on the Sturm-Liouville eigenvalue method. Also, we use the numerical method to study this relation and then compare the results of these methods.

In order to do this, first we need to obtain a solution of Eq. (\ref{z-phi-Eq}) at the critical temperature $T_c$. At $T_c$, the condensation is zero and thus the scalar field vanishes, $\psi=0$. As a result, Eq. (\ref{z-phi-Eq}) leads to
\begin{equation}
\left[1+\frac{bz^4}{r^2_+}\phi'^2(z)\right]\phi''(z)+\frac{2bz^3}{r^2_+}\phi'^3(z)+\frac{4-d}{z}\phi'(z)=0.\label{psizero-phi-Eq}
\end{equation}
The solution of this equation is given as
\begin{equation}
\phi(z)=r_{+c}\int^1_z\frac{\sqrt{W\left(\frac{bC\widetilde{z}^{2(d-2)}}{r^{2(d-2)}_{+c}}\right)}}{\sqrt{b}\widetilde{z}^2}d\widetilde{z},\label{Tc-z-phi-sol}
\end{equation}
where $r_{+c}$ is denoted the event horizon radius of the black hole with the temperature $T_c$, $C$ is an integral constant, and $W(x)$ is the Lambert function defined as \cite{Stegun1972}
\begin{equation}
W(x)e^{W(x)}=x.
\end{equation}
Note that, we have used the boundary condition $\phi(1)=0$. By using the following expansion
\begin{equation}
\sqrt{W(x)}=x^{1/2}-\frac{x^{3/2}}{2}+\mathcal{O}\left(x^{5/2}\right),
\end{equation}
we can expand the solution (\ref{Tc-z-phi-sol}) around $b=0$ as
\begin{equation}
\phi(z)=\frac{\sqrt{C}}{(d-3)r^{d-3}_{+c}}\left(1-z^{d-3}\right)-\frac{bC^{3/2}}{2(3d-7)r^{3d-7}_{+c}}\left(1-z^{3d-7}\right)+\cdots,\label{gau-aysbh}
\end{equation}
From this expansion and the asymptotic behavior of $\phi$ given at (\ref{phi-asy-beh}), we can find $C=\rho^2(d-3)^2$. Near the AdS boundary ($z\rightarrow0$), we express the functions $\phi(z)$ as follows
\begin{equation}
\phi(z)=\lambda r_{+c}\xi(z),\label{near-Ads-phi}
\end{equation}
where $\lambda=\frac{\rho}{r^{d-2}_{+c}}$ and the function $\xi(z)$ is given as
\begin{eqnarray}
\xi(z)&=&\frac{1}{\lambda}\int^1_z\frac{\sqrt{W\left(\frac{bC\widetilde{z}^{2(d-2)}}{r^{2(d-2)}_{+c}}\right)}}{\sqrt{b}\widetilde{z}^2}d\widetilde{z},\nonumber\\
&=&\sqrt{\frac{1}{b\lambda^2}}\int^1_z\frac{\sqrt{W\left(b\lambda^2(d-3)^2\widetilde{z}^{2(d-2)}\right)}}{\widetilde{z}^2}d\widetilde{z}.\label{z-xi-exp}
\end{eqnarray}

In the following, we calculate $b\lambda^2$ by performing the perturbatively iterative procedure. First, let us split the nonlinear parameter $ b$ into the steps as $b_n=n\Delta b$ (for $n=0$, $1$, $2$,...) where $\Delta b=b_{n+1}-b_n$ is the step size which is small for applying the perturbative expansion. And, we denote $\lambda^2_n$, $\lambda^2_{n-1}$, $\lambda^2_{n-2}$,..., $\lambda_0$ to be the values of $\lambda^2$ at $b_n$, $b_{n-1}$, $b_{n-2}$,... , $b_0$, respectively. Then, $b\lambda^2_n$ at the $n$-th step is perturbatively related to $b\lambda^2_{n-1}$ at the $(n-1)$-th step as, $b\lambda^2_n=b\lambda^2_{n-1}+\mathcal{O}\left(\Delta b\right)$. Using the perturbatively iterative procedure, $b\lambda^2$ is computed as follows
\begin{eqnarray}
b\lambda^2&=&b\lambda^2_{n-1}+\mathcal{O}\left(\Delta b\right),\nonumber\\
b\lambda^2_{n-1}&=&b\lambda^2_{n-2}+\mathcal{O}\left(\Delta b\right),\nonumber\\
&\vdots&\nonumber\\
b\lambda^2_{1}&=&b\lambda^2_0+\mathcal{O}\left(\Delta b\right).
\end{eqnarray}
This approximation is more accurate if we consider the smaller step size.

\begin{figure}[t]
 \centering
\begin{tabular}{cc}
\includegraphics[width=0.45 \textwidth]{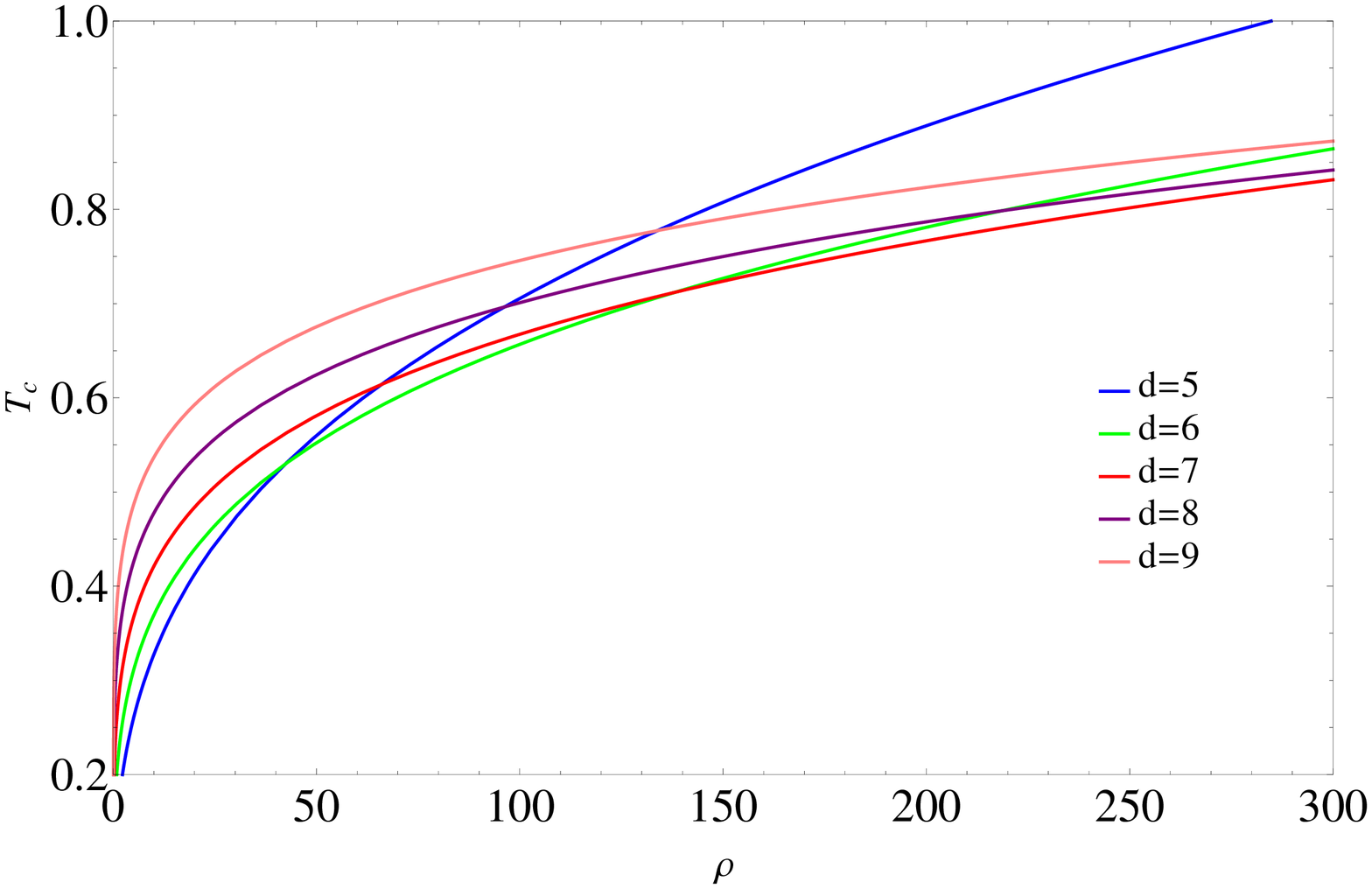}
\hspace*{0.05\textwidth}
\includegraphics[width=0.45 \textwidth]{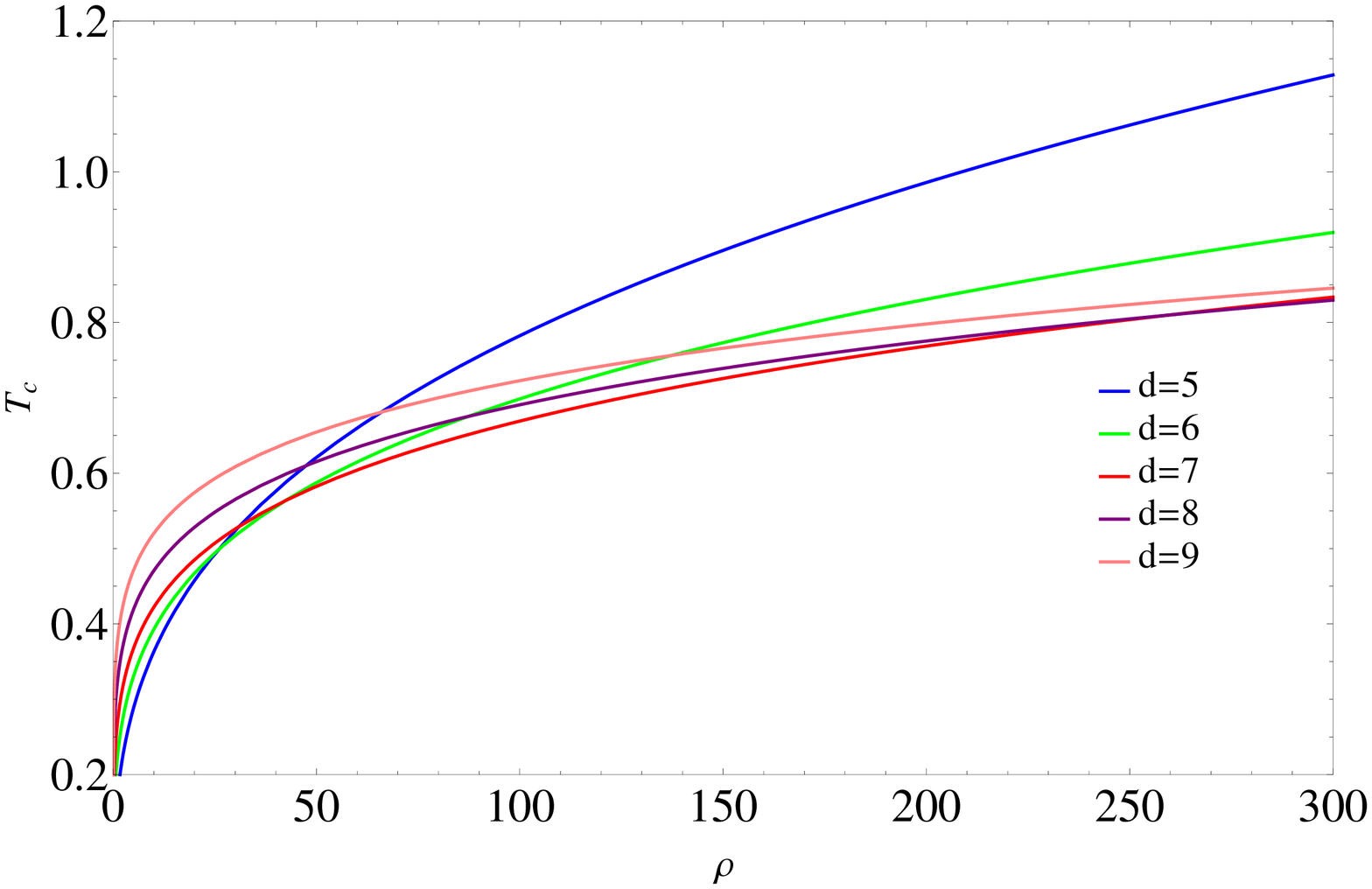}
\end{tabular}
  \caption{Plots of the critical temperature in terms of the charge density, at $b=0.1$. Left panel: $\alpha=0.001$ and $m^2=-2$. Right panel: $\alpha=0.005$ and $m^2=-3$.}\label{fig-Tc-rho}
\end{figure}

With the solution of $\phi(z)$ at the temperature $T_c$, we can find the equation for $\psi$ in the limit $T\rightarrow T_c$ as
\begin{eqnarray}
0&=&\psi''(z)+\left[\frac{2\widetilde{\alpha}(d-5)z^{d-1}-2\left(1-4\widetilde{\alpha}-\sqrt{1-4\widetilde{\alpha}\left(1-z^{d-1}\right)}\right)}{z\sqrt{1-4\widetilde{\alpha}\left(1-z^{d-1}\right)}\left(\sqrt{1-4\widetilde{\alpha}\left(1-z^{d-1}\right)}-1\right)}+\frac{4-d}{z}\right]\psi'(z)\nonumber\\
&&+\left[\frac{4\widetilde{\alpha}^2\lambda^2\xi^2(z)}{\left(\sqrt{1-4\widetilde{\alpha}\left(1-z^{d-1}\right)}-1\right)^2}+\frac{2\widetilde{\alpha}m^2}{z^2\left(\sqrt{1-4\widetilde{\alpha}\left(1-z^{d-1}\right)}-1\right)}\right]\psi(z).\label{nearTc-psi-Eq}
\end{eqnarray}
Near the AdS boundary, we express $\psi(z)$ as follows
\begin{equation}
\psi(z)=\langle\mathcal{O}_+\rangle\frac{z^{\Delta_+}}{r^{\Delta_+}_+}F(z),\label{near-Ads-psi}
\end{equation}
where $F(z)$ is the trial function satisfying the boundary condition $F(0)=1$ and $F'(0)=0$. By substituting Eq. (\ref{near-Ads-psi}) into Eq. (\ref{nearTc-psi-Eq}), we obtain the following equation for $F(z)$
\begin{equation}
F''(z)+p(z)F'(z)+q(z)F(z)+\lambda^2w(z)\xi^2(z)F(z)=0,
\end{equation}
where
\begin{eqnarray}
p(z)&=&\frac{4-d+2\Delta_+}{z}+\frac{2\widetilde{\alpha}(d-5)z^{d-1}-2\left(1-4\widetilde{\alpha}-\sqrt{1-4\widetilde{\alpha}\left(1-z^{d-1}\right)}\right)}{z\sqrt{1-4\widetilde{\alpha}\left(1-z^{d-1}\right)}\left(\sqrt{1-4\widetilde{\alpha}\left(1-z^{d-1}\right)}-1\right)},\nonumber\\
q(z)&=&\frac{\Delta_+(3-d+\Delta_+)}{z^2}+\frac{2\Delta_+\left[\widetilde{\alpha}(d-5)z^{d-1}-\left(1-4\widetilde{\alpha}-\sqrt{1-4\widetilde{\alpha}\left(1-z^{d-1}\right)}\right)\right]}{z^2\sqrt{1-4\widetilde{\alpha}\left(1-z^{d-1}\right)}\left(\sqrt{1-4\widetilde{\alpha}\left(1-z^{d-1}\right)}-1\right)}\nonumber\\
&&+\frac{2\widetilde{\alpha}m^2}{z^2\left(\sqrt{1-4\widetilde{\alpha}\left(1-z^{d-1}\right)}-1\right)},\nonumber\\
w(z)&=&\frac{4\widetilde{\alpha}^2}{\left(\sqrt{1-4\widetilde{\alpha}\left(1-z^{d-1}\right)}-1\right)^2}.
\end{eqnarray}
It is written in the form of the Sturm-Liouville equation as
\begin{equation}
\left[T(z)F'(z)\right]'-Q(z)F(z)+\lambda^2P(z)F(z)=0, \label{SL-eq}
\end{equation}
where
\begin{eqnarray}
Q(z)&=&-T(z)q(z),\nonumber\\
P(z)&=&T(z)w(z)\xi^2(z),\nonumber\\
T(z)&=&e^{\int p(z)dz}=z^{2\Delta_++2-d}\left(1-z^{d-1}\right)\left[1-\widetilde{\alpha}z^{d-1}+\widetilde{\alpha}^2\left(2z^{2d-2}-3z^{d-1}\right)+\mathcal{O}\left(\widetilde{\alpha}^3\right)\right].
\end{eqnarray}
It is known from the Sturm-Liouville eigenvalue problem that the eigenvalues of Eq. (\ref{SL-eq}) are obtained by minimizing the following expression
\begin{equation}
\lambda^2=\frac{\int^1_0T(z)F'^2(z)dz+\int^1_0Q(z)F^2(z)dz}{\int^1_0P(z)F^2(z)dz},\label{Sturm-Liouville}
\end{equation}
where the trial function $F(z)$ is chosen as $F(z)=1-az^2$ \cite{Therrien2010}. For calculating the integrals in (\ref{Sturm-Liouville}), we use the exact form of the function $\xi(z)$ given in Eq. (\ref{z-xi-exp}), and thus the value of the nonlinear parameter $b$ considered is arbitrary without restricting the small value.
With the given parameters, these integrals can be appropriately separated and then integrated numerically. As a result, from Eq. (\ref{Haw-temp}), the critical temperature $T_c$ can be obtained for given nonlinear parameter $b$, GB parameter $\alpha$ and spacetime dimension $d$ as
\begin{equation}
T_c=\frac{(d-1)r_{+c}}{4\pi}=\frac{d-1}{4\pi}\left(\frac{\rho}{\lambda_{\text{min}}}\right)^{\frac{1}{d-2}}\equiv\gamma\rho^{\frac{1}{d-2}},\label{ctemp-chden}
\end{equation}
where $\gamma=\frac{d-1}{4\pi}\lambda_{\text{min}}^{\frac{1}{2-d}}$ is the coefficient of $T_c$. As an example, we determine the coefficient $\gamma$ in the case for $d=5$, $\alpha=0.05$, $b=0.05$, $\Delta b=0.05$, and $m^2=-2$. First, let us calculate $\lambda^2|_{b=0}$ as
\begin{equation}
\lambda^2|_{b=0}=\frac{1.66965-2.55683a+1.23137a^2}{0.0310475-0.0329515a+0.0101823a^2},
\end{equation}
which has a minimum $\lambda^2_{\text{min}}|_{b=0}=36.7815$ at $a=0.784748$. This leads to $b\lambda^2|_{b=0.05}\approx b\lambda^2|_{b=0}=1.83908$. Then, we derive the expression for $\lambda^2|_{b=0.05}$ as
\begin{equation}
\lambda^2|_{b=0.05}=\frac{1.66965-2.55683a+1.23137a^2}{0.0109467-0.0105178a+0.00303046a^2},
\end{equation}
whose minimum is $\lambda^2_{\text{min}}|_{b=0.05}=91.9388$ at $a=0.834335$.
Thus, the expression of the critical temperature given in Eq. (\ref{ctemp-chden}) becomes $T_c=0.149831\rho^{\frac{1}{3}}$. In more detail, we present our analytical result in Table \ref{table-Tc-rho} and Fig. \ref{fig-Tc-rho}.
\begin{table}[!htp]
\centering
\begin{tabular}{|c|c|c|c|c|c|c|c|c|}
  \hline
  \multicolumn{3}{|c|}{$\alpha=0.02$} & \multicolumn{3}{|c|}{$\alpha=0.04$} & \multicolumn{3}{|c|}{$\alpha=0.06$} \\
  \hline
  $b$ & $\text{Analytical}$ & $\text{Numerical}$ & $b$ & $\text{Analytical}$ & $\text{Numerical}$ & $b$ & $\text{Analytical}$ & $\text{Numerical}$\\
  \hline
  0 & $0.194\rho^{1/3}$ & $0.196\rho^{1/3}$ & 0 & $0.191\rho^{1/3}$ & $0.193\rho^{1/3}$ & 0 & $0.188\rho^{1/3}$ & $0.190\rho^{1/3}$ \\ \hline
  0.05 & $0.174\rho^{1/3}$ & $0.168\rho^{1/3}$ & 0.05 & $0.171\rho^{1/3}$ & $0.164\rho^{1/3}$ & 0.05 & $0.168\rho^{1/3}$ & $0.159\rho^{1/3}$ \\ \hline
  0.1 & $0.159\rho^{1/3}$ & $0.153\rho^{1/3}$ & 0.1 & $0.156\rho^{1/3}$ & $0.148\rho^{1/3}$ & 0.1 & $0.152\rho^{1/3}$ & $0.143\rho^{1/3}$ \\ \hline
  0.15 & $0.147\rho^{1/3}$ & $0.141\rho^{1/3}$ & 0.15 & $0.143\rho^{1/3}$ & $0.136\rho^{1/3}$ & 0.15 & $0.139\rho^{1/3}$ & $0.131\rho^{1/3}$ \\ \hline
  \end{tabular}
\caption{A comparison between analytical and numerical values of the critical temperature $T_c$, for the various values of $\alpha$ and $b$, at $d=5$ and $m^2l^2_{\text{eff}}=-3$.}\label{table-Tc-rho}
\end{table}

From Table \ref{table-Tc-rho}, we can see that the critical temperature $T_c$ should decrease as increasing the nonlinear parameter $b$ or the GB parameter $\alpha$, for the other parameters kept fixed. It means that the superconducting phase transition is harder to achieve within the nonlinear electrodynamics as well as Einstein-Gauss-Bonnet gravity. This can been seen more clearly when we study condensation value in later. Furthermore, from Fig. \ref{fig-Tc-rho}, we observe that the superconducting phase transition in the higher (lower) dimensional systems can achieve at the higher critical temperature at the efficiently low (high) charge density.

Now we introduce the numerical study relying on the shooting method to check the result obtained by the analytic or Sturm-Liouville method. First, it should be noted that Eqs. (\ref{z-phi-Eq}) and (\ref{z-psi-Eq})
can be written as
\begin{eqnarray}
 \left[1+bz^4\bar{\phi}'^2(z)\right]\bar{\phi}''(z)+2bz^3\bar{\phi}'^3(z)+\frac{4-d}{z}\bar{\phi}'(z)-\frac{2\bar{\phi}(z)}{z^4h(z)}\psi^2(z)e^{-\frac{bz^4}{2}\bar{\phi}'^2(z)}&=&0,\label{z-phi-Eq3}\\
 \psi''(z)+\left(\frac{h'(z)}{h(z)}+\frac{4-d}{z}\right)\psi'(z)+\frac{1}{z^4}\left(\frac{\bar{\phi}^2(z)}{h^2(z)}-\frac{m^2}{h(z)}\right)\psi(z)&=&0,\label{z-psi-Eq4}
\end{eqnarray}
where $\bar{\phi}(z)\equiv\phi(z)/r_+$ and $h(z)=f(z)/r^2_+$. For obtaining the initial values, we use the behavior of $\bar{\phi}$ and $\psi$ near the horizon ($z\rightarrow1$), given as
\begin{eqnarray}
\bar{\phi}(z)&=&\bar{\phi}'(1)(z-1)+\frac{\bar{\phi}''(1)}{2}(z-1)^2+\frac{\bar{\phi}'''(1)}{3!}(z-1)^3+\cdots,\\
\psi(z)&=&\psi(1)+\psi'(1)(z-1)+\frac{\psi''(1)}{2}(z-1)^2+\frac{\psi'''(1)}{3!}(z-1)^3+\cdots.
\end{eqnarray}
By replacing these expansion into Eqs. (\ref{z-phi-Eq3}) and (\ref{z-psi-Eq4}), we can calculate the coefficients $\bar{\phi}''(1)$, $\bar{\phi}'''(1)$,... and $\psi'(1)$, $\psi''(1)$, $\psi'''(1)$,... in terms of $\bar{\phi}'(1)$ and $\psi(1)$. Note that, the higher order terms are very small and thus can be neglected.
Near the critical temperature, the value of $\psi$ is very small and thus we can set $\psi(1)=10^{-6}$. Using the shooting method, we should solve numerically Eqs. (\ref{z-phi-Eq3}) and (\ref{z-psi-Eq4}), with the given parameters, near the horizon with the shooting parameter $\bar{\phi}'(1)$ to get proper solutions at the AdS boundary ($z=0$). Here, the behavior of $\bar{\phi}$ and $\psi$ near the AdS boundary is given as
\begin{eqnarray}
\bar{\phi}(z)&=&\frac{\mu}{r_+}-\frac{\rho}{r^{d-2}_+}z^{d-3},\\
\psi(z)&=&\frac{\langle\mathcal{O}_-\rangle}{r^{\Delta_-}_+}z^{\Delta_-}+\frac{\langle\mathcal{O}_+\rangle}{r^{\Delta_+}_+}z^{\Delta_+}.
\end{eqnarray}
For the specific value of $\bar{\phi}'(1)$, we impose the boundary condition $\langle\mathcal{O}_-\rangle/r^{\Delta_-}_+=0$. With this value of $\bar{\phi}'(1)$, we can specify the coefficients in the asymptotic behavior of and $\bar{\phi}$ and $\psi$. From the specified value of $\frac{\rho}{r^{d-2}_{+c}}$, we can find the relation between the critical temperature and the charge density. The numerical values are compared to the analytic values in Table \ref{table-Tc-rho}. Here, we can see that both results are in good agreement with together. It should be noted that in the calculations we only stop up to the second-order in the GB parameter and the step size is $\Delta b=0.05$. If the higher-order terms in the GB parameter are taken into account and the step size $\Delta b$ is smaller, it is expected to reduce the disparity between the analytic and numerical methods.

\section{\label{CVCE} Condensation value and critical exponent}
In this section, we will calculate the condensation value $\langle\mathcal{O}_+\rangle$ and the critical exponent for GB holographic superconductors in the exponential nonlinear electrodynamics. Thus, we need to study the behavior of the gauge field near the critical temperature $T_c$. The equation describes the behavior of the gauge field near $T_c$ up to second-order in the nonlinear parameter $b$ as
\begin{eqnarray}\label{nearTc-ph-Eq}
0&=&\phi''(z)+\frac{4-d}{z}\phi'(z)+\frac{(d-2)bz^3}{r^2_+}\phi'^3(z)+\frac{(2-d)b^2z^7}{r^4_+}\phi'^5(z)\nonumber\\
&&-\frac{2r^2_+\phi(z)\psi^2(z)}{z^4f(z)}\left[1-\frac{3bz^4}{2r^2_+}\phi'^2(z)+\frac{13b^2z^8}{8r^4_+}\phi'^4(z)\right]+\mathcal{O}\left(b^3\right).
\end{eqnarray}
(It should be noted that the equation of the gauge field for Born-Infeld and exponential nonlinear electrodynamics are same together up to first-order in the nonlinear parameter $b$.) Near the critical temperature $T_c$, the condensation value $\langle\mathcal{O}_+\rangle$ is very small. Since we can expand the $\phi(z)$ in $\langle\mathcal{O}_+\rangle$ in the vicinity of the AdS boundary as
\begin{equation}
\frac{\phi(z)}{r_+}=\lambda\xi(z)+\frac{\langle\mathcal{O}_+\rangle^2}{r^{2\Delta_+}_+}\chi(z)+\cdots,\label{phi-Exp}
\end{equation}
where the function $\chi(z)$ satisfies the boundary condition, $\chi(1)=\chi'(1)=0$. By substituting this expansion into Eq. (\ref{nearTc-ph-Eq}), we find
\begin{eqnarray}
0&=&\chi''(z)+\left[\frac{4-d}{z}+3(d-2)b\lambda^2z^3\xi'^2(z)+5(2-d)b^2\lambda^4z^7\xi'^4(z)\right]\chi'(z)\nonumber\\
&&-\frac{2\lambda r^2_+\xi(z)}{f(z)}z^{-4+2\Delta_+}F^2(z)\left[1-\frac{3b\lambda^2}{2}z^4\xi'^2(z)+\frac{13
b^2\lambda^4}{8}z^8\xi'^4(z)\right]=0.\label{chi-Eq}
\end{eqnarray}
Using Eq. (\ref{psizero-phi-Eq}) whose solution is given at (\ref{near-Ads-phi}), we find
\begin{equation}
b\lambda^2z^4\xi''(z)\xi'(z)=(d-4)b\lambda^2z^3\xi'^2(z)+(2-d)b^2\lambda^4z^7\xi'^4(z)+\mathcal{O}\left(b^3\right).
\end{equation}
With this relation, Eq. (\ref{chi-Eq}) can be written up to second-order in the nonlinear parameter $b$ as
\begin{equation}
\left[g(z)\chi'(z)\right]'=\frac{2\lambda r^2_+\xi(z)}{f(z)}z^{-d+2\Delta_+}F^2(z),
\end{equation}
where the function $g(z)$ is given as
\begin{equation}
g(z)=z^{4-d}e^{\frac{b\lambda^2z^4\xi'^2(z)}{2}\left[3-b\lambda^2z^4\xi'^2(z)\right]}.
\end{equation}
Then, by integrating the above equation with using the condition $\chi'(1)=0$, we obtain
\begin{eqnarray}
\chi'(z)&=&z^{d-4}e^{\frac{b\lambda^2z^4\xi'^2(z)}{2}\left[3-b\lambda^2z^4\xi'^2(z)\right]}\int^{z}_1\frac{2\lambda r^2_+\xi(\widetilde{z})}{f(\widetilde{z})}\widetilde{z}^{-d+2\Delta_+}F^2(\widetilde{z})d\widetilde{z}.\label{Der-psi}
\end{eqnarray}

From Eqs. (\ref{phi-asy-beh}) and (\ref{phi-Exp}), we get the relation
\begin{equation}
\frac{\mu}{r_+}-\frac{\rho}{r^{d-2}_+}z^{d-3}=\lambda\xi(z)+\frac{\langle\mathcal{O}_+\rangle^2}{r^{2\Delta_+}_+}\left[\chi(0)+\chi'(0)z+\frac{\chi''(0)}{2}z^2+\cdots+\frac{\chi^{(d-3)}(0)}{(d-3)!}z^{d-3}+\cdots\right].
\end{equation}
By comparing the coefficients of the terms relating to $z^{d-3}$ in the right-hand and left-hand sides of the above equation, we find
\begin{equation}
\frac{\rho}{r^{d-2}_+}=\lambda-\frac{\langle\mathcal{O}_+\rangle^2}{r^{2\Delta_+}_+}\frac{\chi^{(d-3)}(0)}{(d-3)!}.
\end{equation}
[Also, we can determine $\chi'(0)=\chi''(0)=\cdots=\chi^{(d-4)}(0)=0$ which is consistent to the expression of $\chi'(z)$ given in Eq. (\ref{Der-psi}).] From this, we find the expression for the order parameter $\langle\mathcal{O}_+\rangle$ as
\begin{eqnarray}
\langle\mathcal{O}_+\rangle=\frac{\beta  T^{\Delta_+}}{\sqrt{d-2}}\sqrt{\left(\frac{T_c}{T}\right)^{d-2}-1}\approx\beta  T^{\Delta_+}_c\sqrt{1-\frac{T}{T_c}},
\end{eqnarray}
where
\begin{equation}
\beta=\left(\frac{4\pi}{d-1}\right)^{\Delta_+}\sqrt{-\frac{\lambda(d-2)!}{\chi^{(d-3)}(0)}}.
\end{equation}
This expression indicates that the critical exponent is $1/2$ which is universal independently on the nonlinear electrodynamics and GB corrections.

In order to calculate $\beta$, we use a fact that in the limit $z\rightarrow0$ Eq. (\ref{chi-Eq}) becomes
\begin{equation}
\chi''(0)=\frac{d-4}{z}\chi'(z)\Big|_{z\rightarrow0}.
\end{equation}
This leads to the following relation between the $n$-th and first-order derivatives of $\chi$ at $z=0$
\begin{equation}
\chi^{(n)}(0)=(d-4)(d-5)...(d-2-n)\frac{\chi'(z)}{z^{n-1}}\Big|_{z\rightarrow0}.
\end{equation}
Using this relation with $\chi'(z)$ given at Eq. (\ref{Der-psi}), we find $\beta$ as
\begin{equation}
\beta=\left(\frac{4\pi}{d-1}\right)^{\Delta_+}\sqrt{\frac{(d-2)(d-3)}{\mathcal{A}}},
\end{equation}
where $\mathcal{A}$ is given by
\begin{eqnarray}
\mathcal{A}&=&\int^{1}_0\frac{2r^2_+\xi(z)}{f(z)}z^{-d+2\Delta_+}F^2(z)dz,\nonumber\\
&=&4\widetilde{\alpha}\sqrt{\frac{1}{b\lambda^2}}\int^{1}_0\frac{z^{2(1+\Delta_+)-d}(1-az^2)^2}{1-\sqrt{1-4\widetilde{\alpha}(1-z^{d-1})}}\left[\int^1_z\frac{\sqrt{W\left(b\lambda^2(d-3)^2\widetilde{z}^{2(d-2)}\right)}}{\widetilde{z}^2}d\widetilde{z}\right]dz,
\end{eqnarray}
which can be, with the given parameters, integrated numerically.

\begin{figure}[t]
 \centering
\begin{tabular}{cc}
\includegraphics[width=0.45 \textwidth]{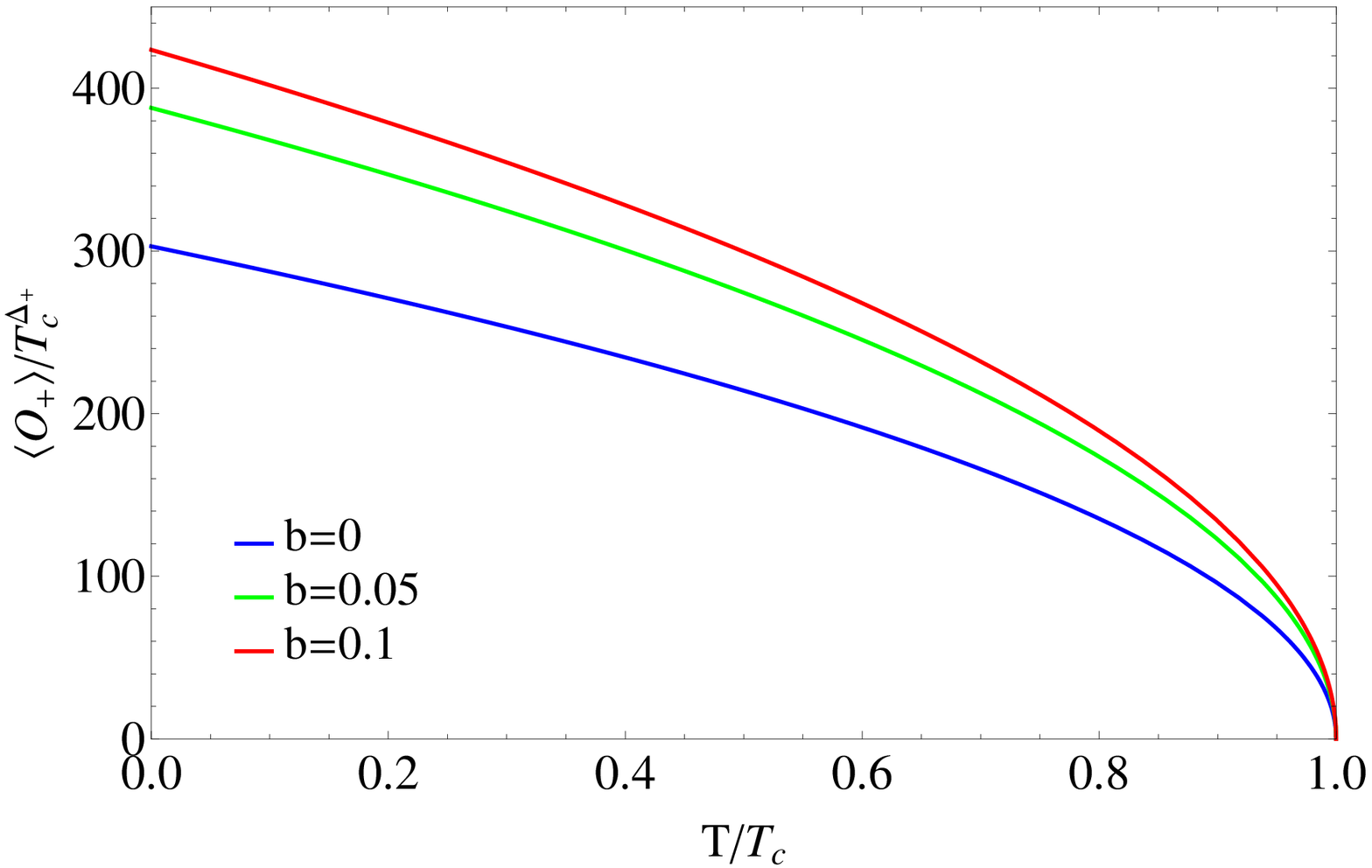}
\hspace*{0.05\textwidth}
\includegraphics[width=0.45 \textwidth]{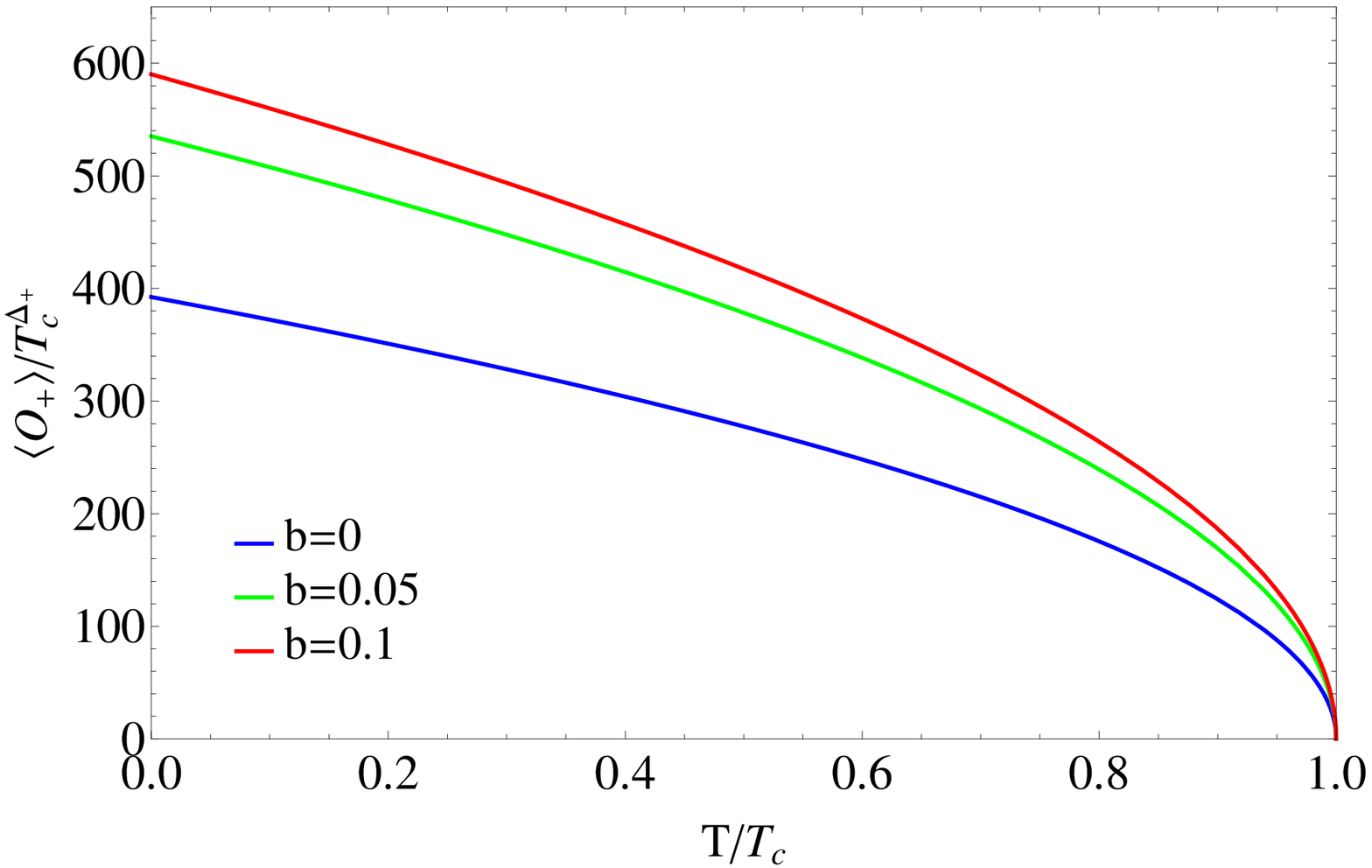}
\end{tabular}
  \caption{The dimensionless condensation operator as a function of the dimensionless temperature for the different values of $b$, at $d=5$ and $m^2=-3$. Left panel: $\alpha=0.04$. Right panel: $\alpha=0.08$.}\label{orderpaer-temp-1}
\end{figure}

\begin{figure}[t]
 \centering
\begin{tabular}{cc}
\includegraphics[width=0.45 \textwidth]{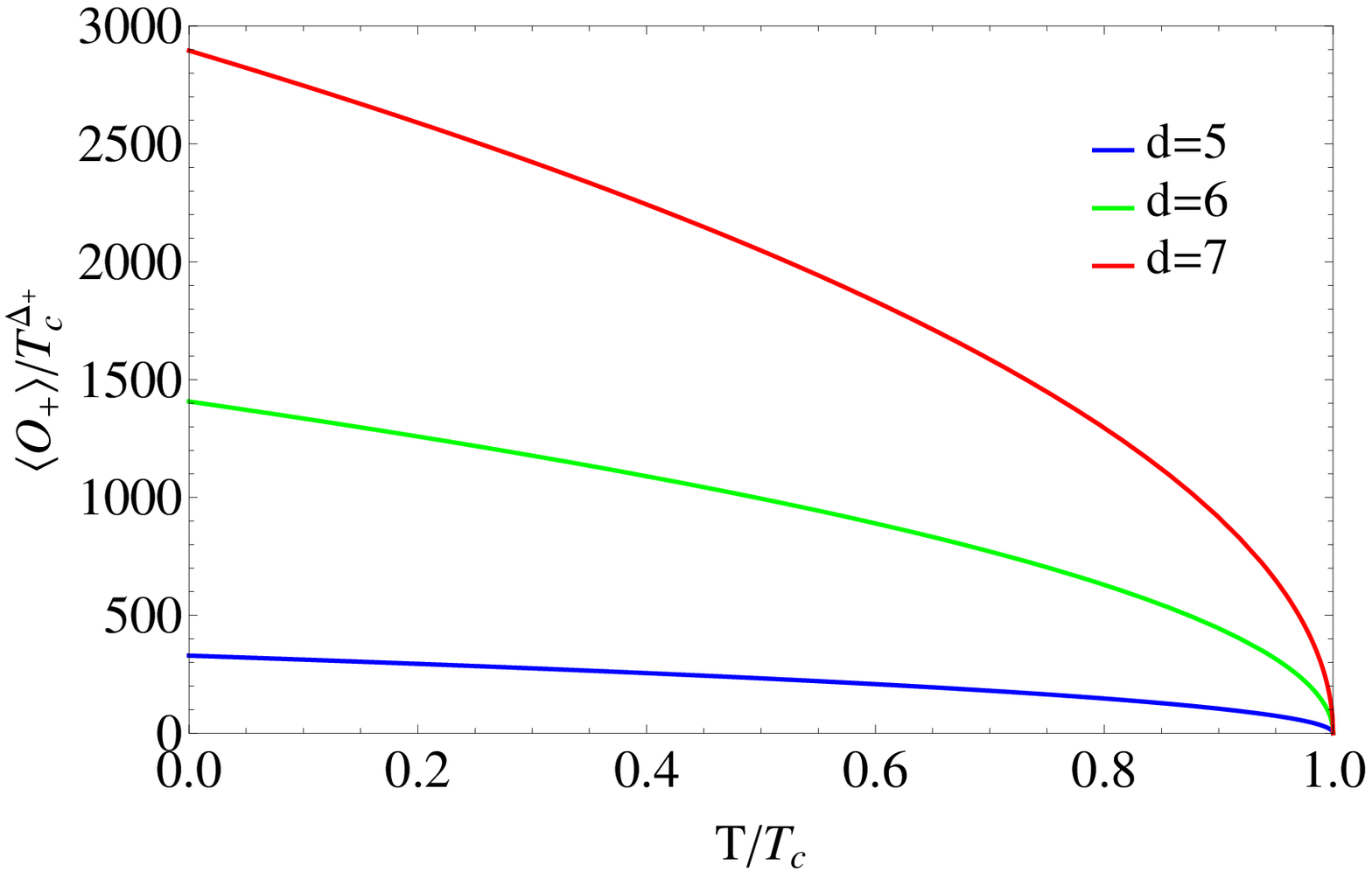}
\hspace*{0.05\textwidth}
\includegraphics[width=0.45 \textwidth]{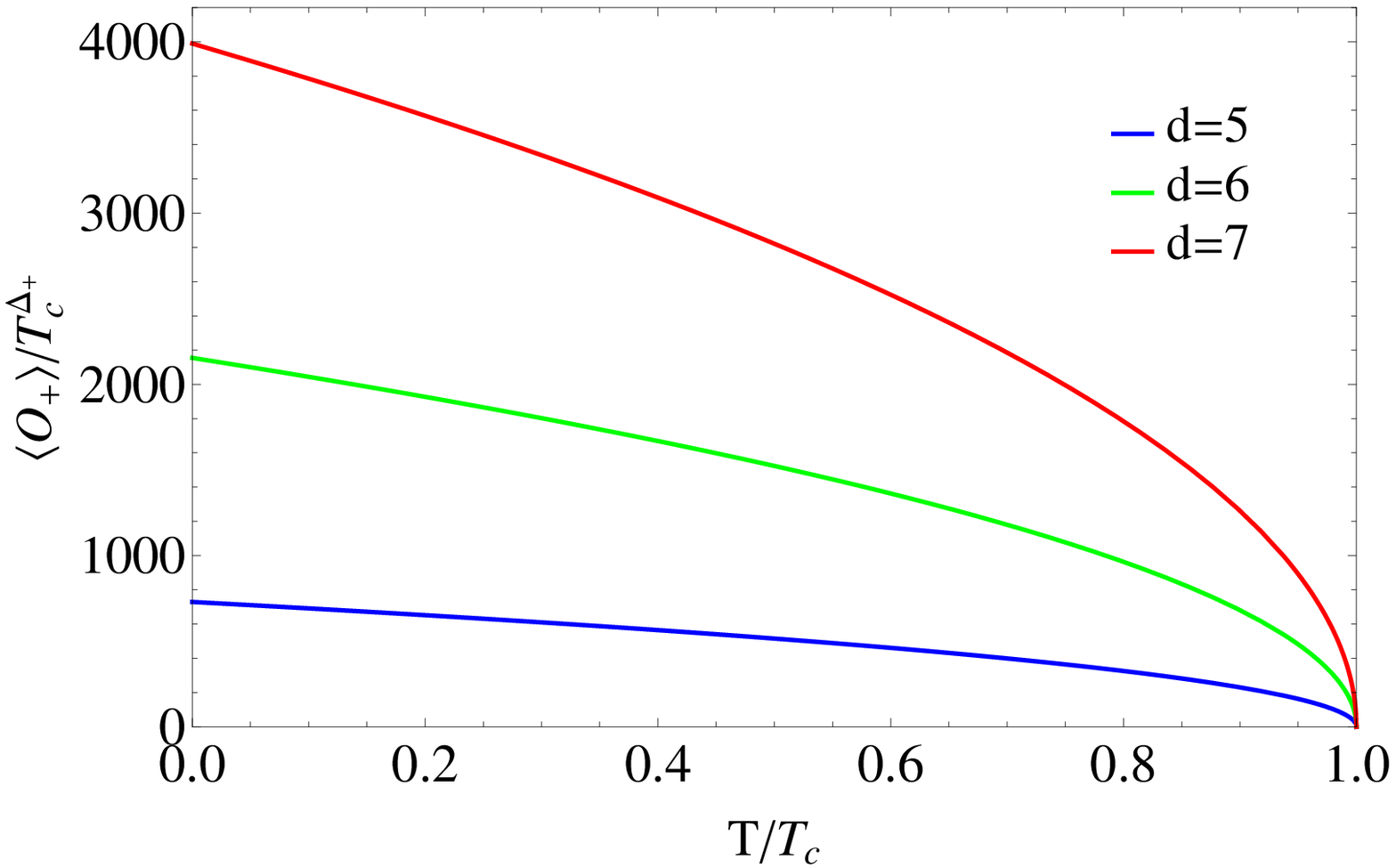}
\end{tabular}
  \caption{The dimensionless condensation operator as a function of the dimensionless temperature for various dimension, at $b=0.1$. Left panel: $\alpha=0.005$ and $m^2=-3$. Right panel: $\alpha=0.01$ and $m^2=-2$.}\label{orderpaer-temp-2}
\end{figure}

Effects of the nonlinear electrodynamics, the GB corrections and spacetime dimension on the condensation value are explicitly shown in Figs. \ref{orderpaer-temp-1} and \ref{orderpaer-temp-2}. These figures indicate that, for $T/T_c$ kept fixed, the condensation value becomes larger as increasing the nonlinear parameter, the GB parameter, or the spacetime dimension, with the other parameters kept fixed. It means that the nonlinear electrodynamics, the GB correction, or the higher sapcetime dimension make the condensation harder, which is consistent to what we obtained in the previous section.

\section{\label{HC} Holographic conductivity}

In this section, we will study the optical conductivity (the conductivity as a function of the frequency) and superconducting energy gap for GB holographic superconductors in the exponential nonlinear electrodynamics. According to AdS/CFT correspondence, the fluctuations of the vector gauge field in the bulk theory gives rise the conductivity in the boundary theory. Thus, in order to compute the conductivity, we should turn on, for simplicity, the small perturbation of the component $A_x$ in the background of the bulk gauge potential as
\begin{equation}
A_x=A_x(r)e^{-i\omega t},
\end{equation}
with $\omega$ to be the frequency. Then, the equation for $A_x(r)$ is obtained up to the linear order in $A_x(r)$ as
\begin{equation}
A''_x(r)+\left[\frac{d-4}{r}+\frac{f'(r)}{f(r)}+b\phi'(r)\phi''(r)\right]A'_x(r)+\frac{\omega^2}{f^2(r)}A_x(r)-\frac{2\psi^2(r)}{f(r)}A_x(r)e^{-\frac{b}{2}\phi'^2(r)}=0.\label{Ax-EOM}
\end{equation}
In the asymptotic limit ($r\rightarrow\infty$), this equation becomes
\begin{equation}
A''_x(r)+\frac{d-2}{r}A'_x(r)+\frac{\omega^2l^4_{\text{eff}}}{r^4}A_x(r)=0,
\end{equation}
whose solution is given by
\begin{equation}\label{aaa}
A_{x}=\left\{
\begin{array}{rl}
A^{(0)}_x+\frac{A^{(1)}_x}{r^{2}}+\frac{\omega ^{2}l^4_{\text{eff}}A^{(0)}_x\ln(\Lambda r)}{2r^{2}}+\cdots , &
\ \quad \mathrm{for}\ d=5, \\
& \\
A^{(0)}_x+\frac{A^{(1)}_x}{r^3}+\frac{\omega ^{2}l^4_{\text{eff}}A^{(0)}_x}{2r^{2}}+\cdots \ , & \ \ ~~%
\mathrm{for}\ d=6, \\
& \\
A^{(0)}_x+\frac{A^{(1)}_x}{r^4}+\frac{\omega^2l^4_{\text{eff}}A^{(0)}_x}{r^2}+\frac{\omega^{4}l^8_{\text{eff}}A^{(0)}_x\ln(\Lambda r)}{16r^{4}}+\cdots \ , & \ \ ~~%
\mathrm{for}\ d=7, \\
& \\
\end{array}%
\right.
\end{equation}%
where $\Lambda$ is a constant with the dimension of the inverse length.

According to the AdS/CFT correspondence, the expectation value of the current operator in the dual field theory on the boundary is determined as \cite{Horowitz2008a,Horowitz2008b}
\begin{equation}
\langle J_x\rangle=\frac{\delta S_{\text{o.s}}}{\delta A^{(0)}_x},
\end{equation}
where $S_{\text{o.s}}$ is the on-shell action given by
\begin{equation}
S_{\text{o.s}}=\int d^{d-1}x\int^{\infty}_{r_+}dr\sqrt{-g}\mathcal{L}_{\text{m}},
\end{equation}
which is calculated on the equations of motion. In the approximation to quadratic order with respect to the pertubation of the gauge field component $A_x$ and up to the irrelevant terms, we have
\begin{equation}
S_{\text{o.s}}=\int d^{d-1}x\int^{\infty}_{r_+}drr^{d-2}\left[-e^{\frac{b\phi'(r)^2}{2}}\left(\frac{\omega^2}{2r^2f(r)}A^2_x+\frac{f(r)}{2r^2}A'^2_x\right)-\frac{\psi^2}{r^2}A^2_x\right].
\end{equation}
Using Eq. (\ref{Ax-EOM}), the on-shell action is rewritten as
\begin{eqnarray}
S_{\text{o.s}}&=&-\frac{1}{2}\int d^{d-1}x\int^{\infty}_{r_+}dre^{\frac{b\phi'(r)^2}{2}}\left\{\left[(d-4)r^{d-5}f(r)+r^{d-4}f'(r)+r^{d-4}f(r)b\phi'(r)\phi''(r)\right]A'_xA_x+\right.\nonumber\\
&&\left.+r^{d-4}f(r)A'^2_x+r^{d-4}f(r)A''_xA_x\right\},\nonumber\\
&=&-\frac{1}{2}\int d^{d-1}x\left[e^{\frac{b\phi'(r)^2}{2}}r^{d-4}f(r)A'_xA_x\right]\Big|_{r\rightarrow\infty}.
\end{eqnarray}
It is important to note here that in the above expression there are the appearance of divergences, such as the logarithmic divergence in the case of five dimensions. However, these divergences are removed by the holographic renormalization by adding a suitable counterterm at the boundary. For constructing the boundary counterterm action, we use Skenderis's holographic renormalization method \cite{Skenderis2002,Sutcliffe2010,Wu-Zhang2013}. More explicitly, the boundary counterterm action is given as
\begin{eqnarray}
S_{\text{c.t.}}=\left\{
\begin{array}{rl}
-\frac{l^2_{\text{eff}}}{2}\ln\left(\frac{\epsilon}{\Lambda}\right)\Large{\int} d^4xA_i\partial^2_tA_i\Big|_{r=\epsilon^{-1}}, &
\ \quad \mathrm{for}\ d=5, \\
& \\
\frac{l^2_{\text{eff}}}{2\epsilon}\Large{\int} d^5xA_i\partial^2_tA_i\Big|_{r=\epsilon^{-1}}, & \ \ ~~%
\mathrm{for}\ d=6, \\
& \\
\frac{l^6_{\text{eff}}}{8}\ln\left(\frac{\epsilon}{\Lambda}\right)\Large{\int} d^6xA_i\partial^4_tA_i\Big|_{r=\epsilon^{-1}}+\frac{l^2_{\text{eff}}}{\epsilon^2}\Large{\int} d^6xA_i\partial^2_tA_i\Big|_{r=\epsilon^{-1}}, & \ \ ~~%
\mathrm{for}\ d=7, \\
& \\
\end{array}%
\right.
\end{eqnarray}%
where $\epsilon^{-1}$ is the cutoff length. With this result, we derive the expectation value of the boundary current operator as
\begin{equation}\label{aaa}
\langle J_x\rangle=\left\{
\begin{array}{rl}
\frac{A^{(1)}_x}{l^2_{\text{eff}}}-\frac{\omega^{2}l^2_{\text{eff}}A^{(0)}_x}{2}, &
\ \quad \mathrm{for}\ d=5, \\
& \\
\frac{3A^{(1)}_x}{2l^2_{\text{eff}}}, & \ \ ~~%
\mathrm{for}\ d=6, \\
& \\
\frac{2A^{(1)}_x}{l^2_{\text{eff}}}+\frac{31\omega^{4}l^6_{\text{eff}}A^{(0)}_x}{16}, & \ \ ~~%
\mathrm{for}\ d=7. \\
& \\
\end{array}%
\right.
\end{equation}%

With $E_x=-\partial_tA_x=i\omega A_x$, we obtain the electric conductivity as
\begin{equation}\label{aaa}
\sigma=\frac{\langle J_x\rangle}{E_x}=\left\{
\begin{array}{rl}
\frac{A^{(1)}_x}{i\omega l^2_{\text{eff}}A^{(0)}_x}+\frac{i\omega l^2_{\text{eff}}}{2}, &
\ \quad \mathrm{for}\ d=5, \\
& \\
\frac{3A^{(1)}_x}{2i\omega l^2_{\text{eff}}A^{(0)}_x}, & \ \ ~~%
\mathrm{for}\ d=6, \\
& \\
\frac{2A^{(1)}_x}{i\omega l^2_{\text{eff}}A^{(0)}_x}-\frac{31i\omega^{3}l^6_{\text{eff}}}{16}, & \ \ ~~%
\mathrm{for}\ d=7. \\
& \\
\end{array}%
\right.
\end{equation}
We observe that the electric conductivity is directly dependent on the GB coupling and the spacetime dimension. The nonlinear parameter $b$ affects indirectly the electric conductivity via $A^{(0)}_x$ and $A^{(1)}_x$.
With this expression, we can calculate the electric conductivity by solving numerically Eq. (\ref{Ax-EOM}). Let us rewrite this equation in the coordinate $z$ as
\begin{eqnarray}
0&=&A''_x(z)+\left[\frac{6-d}{z}+\frac{f'(z)}{f(z)}+bz^3\phi'(z)\left[z\phi''(z)+2\phi'(z)\right]\right]A'_x(z)+\frac{\omega^2}{z^4f^2(z)}A_x(z)\nonumber\\
&&-\frac{2\psi^2(z)}{z^4f(z)}A_x(z)e^{-\frac{bz^4}{2}\phi'^2(z)},\label{z-Ax-EOM}
\end{eqnarray}
and solve it with an ingoing wave boundary condition
\begin{equation}
A_x(z)=f(z)^{-\frac{i\omega}{d-1}}\left[1+a_1(z-1)+a_2(z-1)^2+\cdots\right],
\end{equation}
where $a_1$, $a_2$,... are the coefficients of the Taylor expansion.

\begin{figure}[t]
 \centering
\begin{tabular}{cc}
\includegraphics[width=0.45 \textwidth]{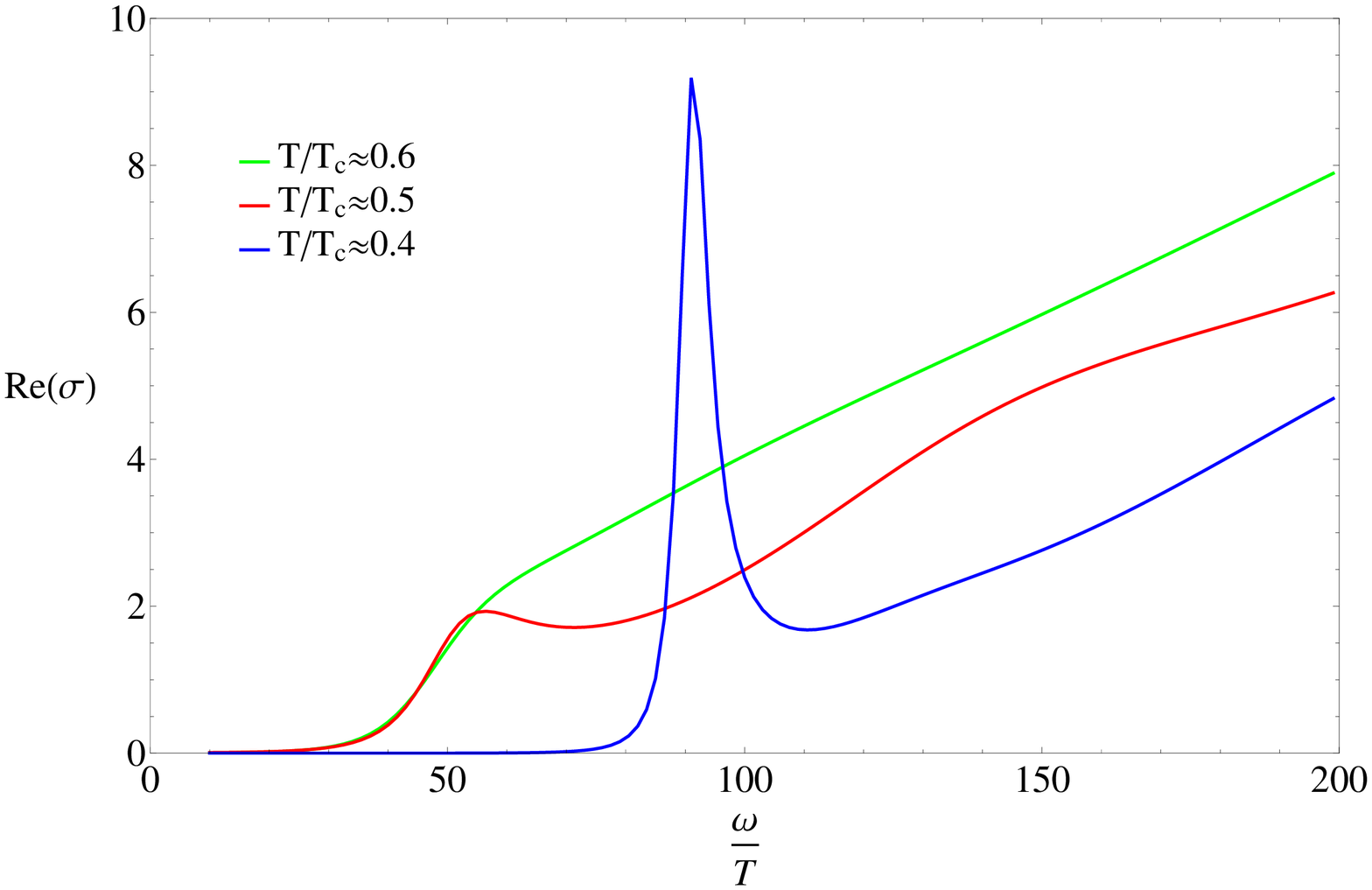}
\hspace*{0.05\textwidth}
\includegraphics[width=0.45 \textwidth]{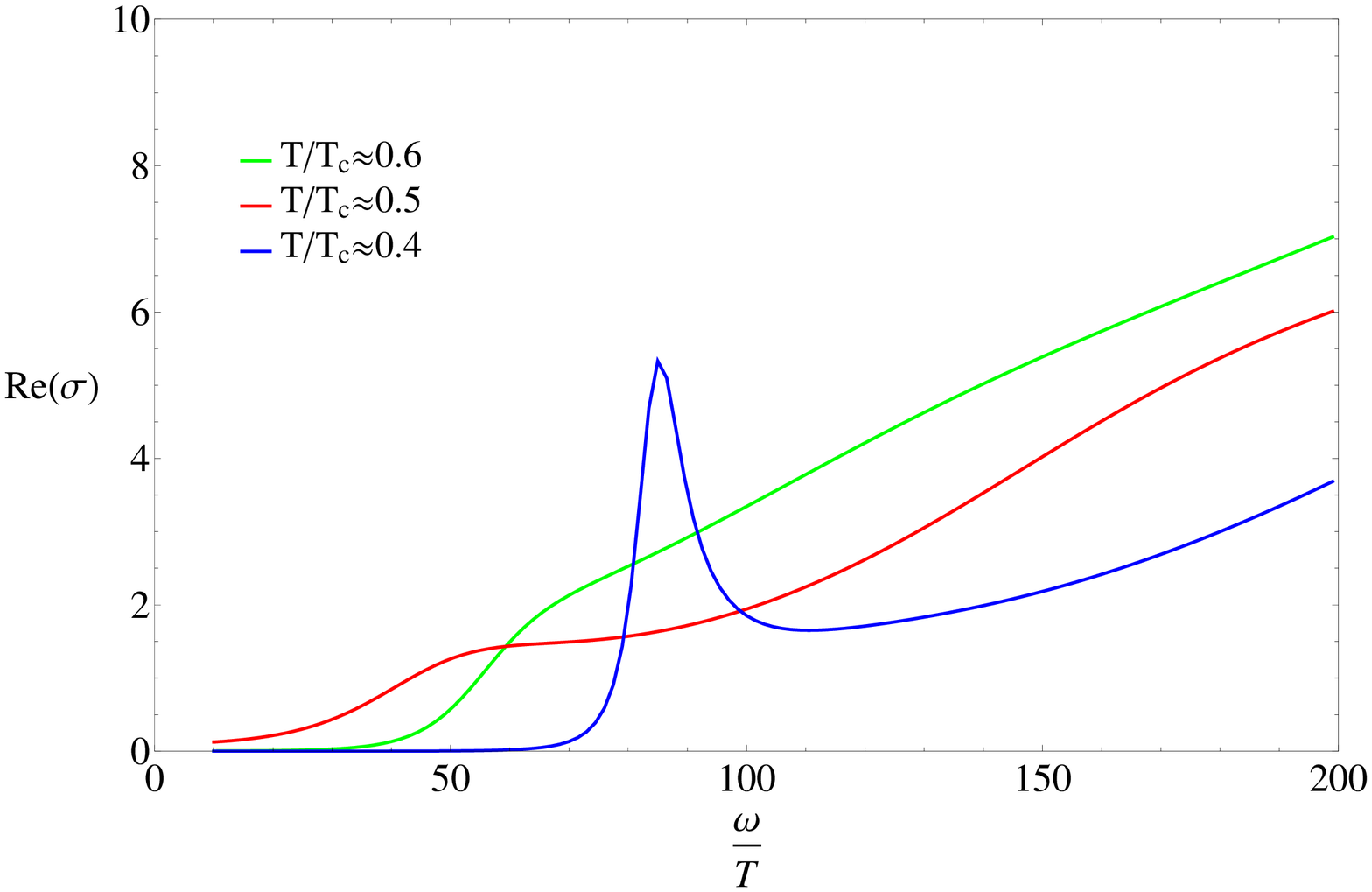}
\end{tabular}
  \caption{The real part of the conductivity as a function in terms of $\omega/T$ for the different values of $T/T_c$ and $b$, at $d=5$, $\alpha=0.06$ and  $m^2l^2_{\text{eff}}=-3$. Left panel: $b=0.05$. Right panel: $b=0.1$.}\label{Re-conduc-d=5-alpha=0.06}
\end{figure}

\begin{figure}[t]
 \centering
\begin{tabular}{cc}
\includegraphics[width=0.44 \textwidth]{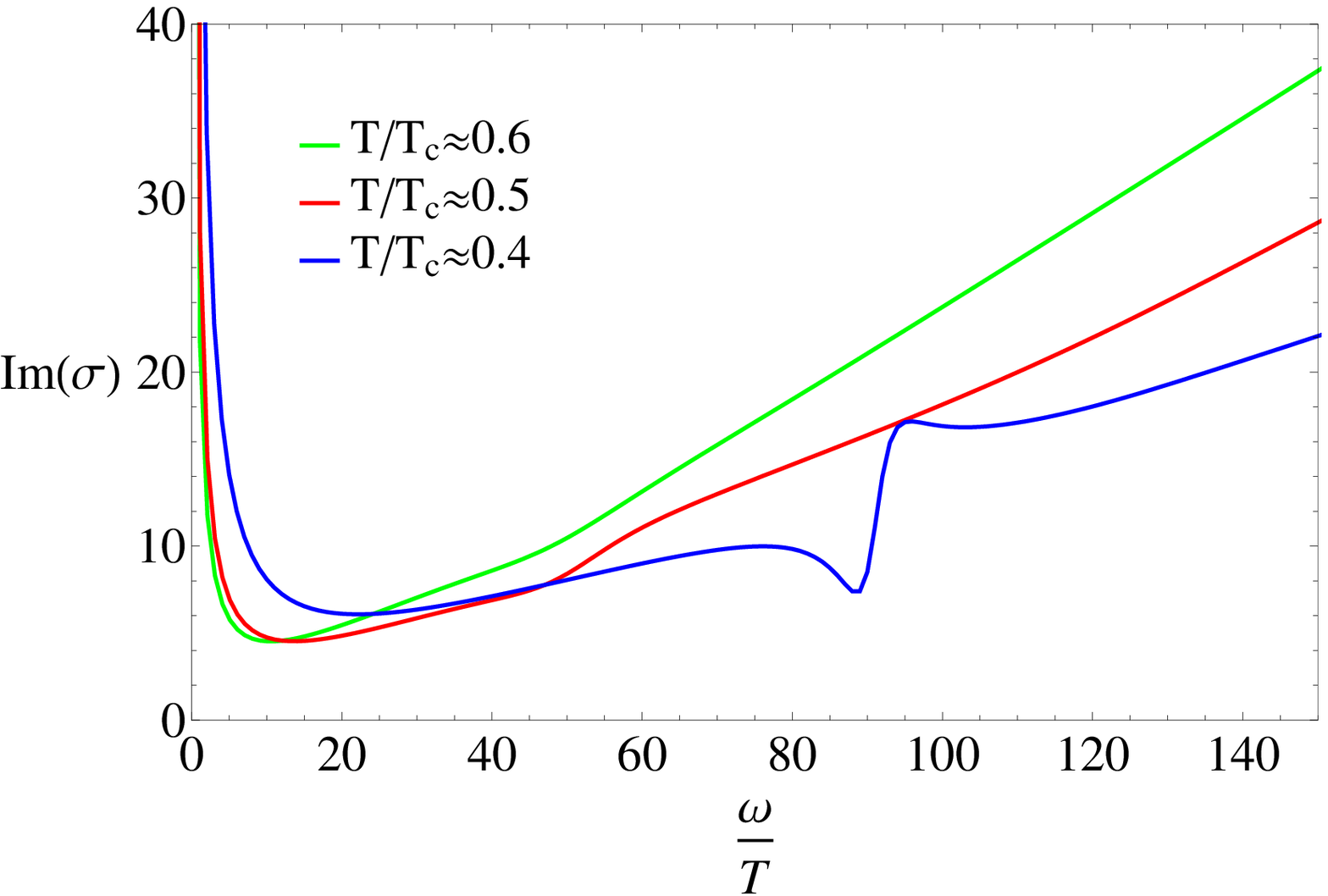}
\hspace*{0.05\textwidth}
\includegraphics[width=0.45 \textwidth]{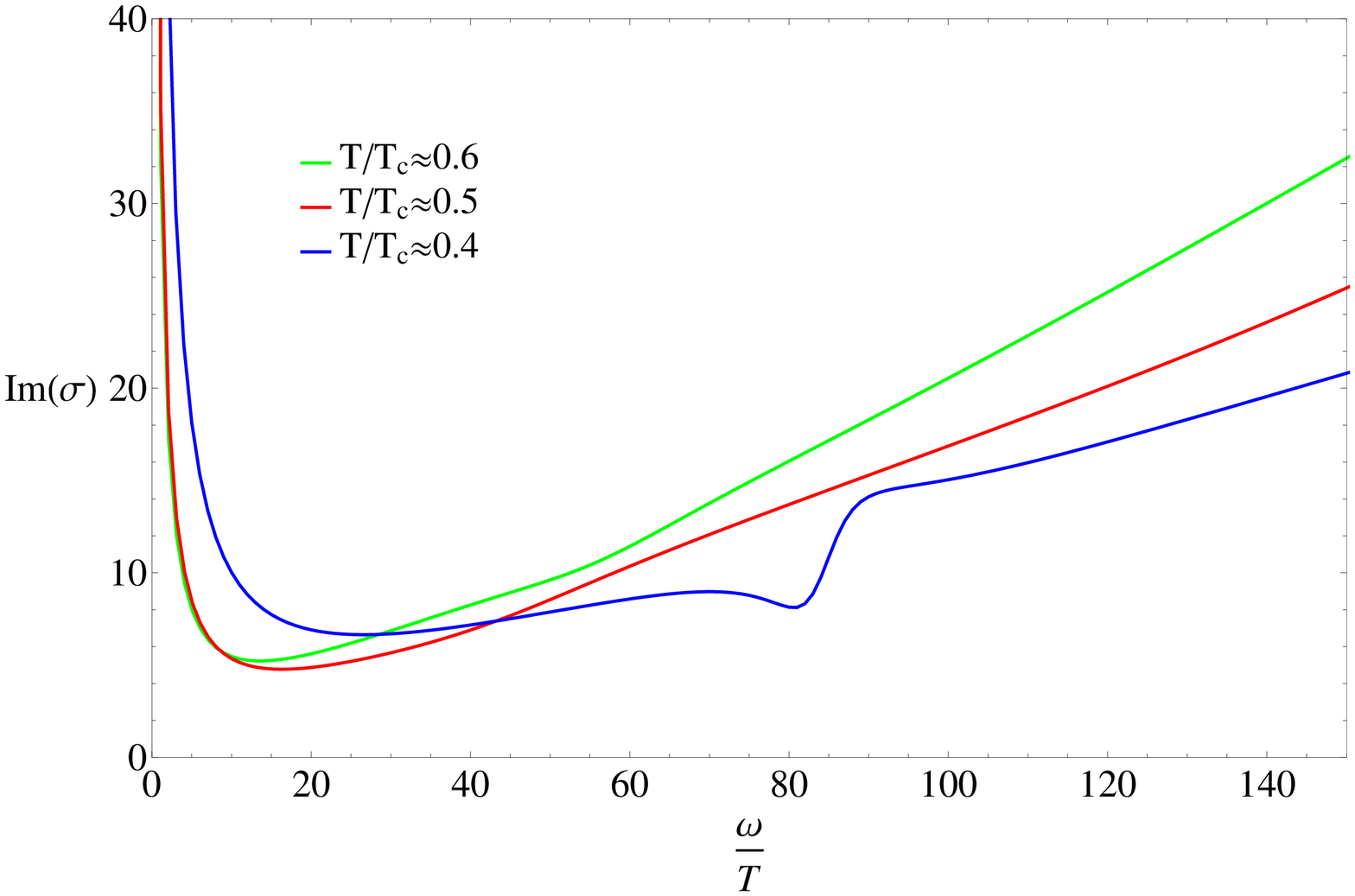}
\end{tabular}
  \caption{The behavior of the imaginary part of the conductivity in terms of $\omega/T$ for the different values of $T/T_c$ and $b$, at $d=5$, $\alpha=0.06$ and  $m^2l^2_{\text{eff}}=-3$. Left panel: $b=0.05$. Right panel: $b=0.1$.}\label{Im-conduc-d=5-alpha=0.06}
\end{figure}

\begin{figure}[t]
 \centering
\begin{tabular}{cc}
\includegraphics[width=0.45 \textwidth]{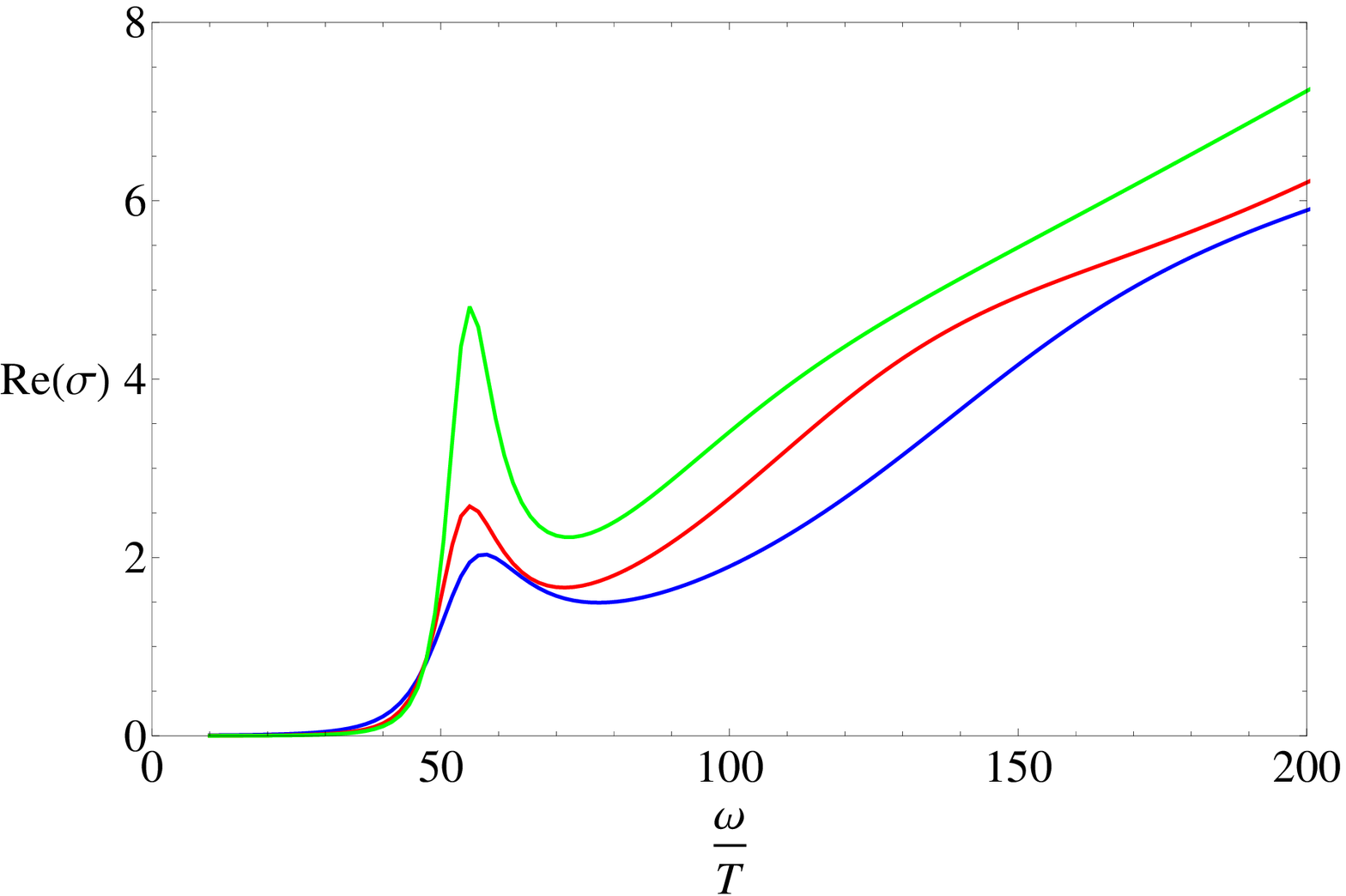}
\hspace*{0.05\textwidth}
\includegraphics[width=0.45 \textwidth]{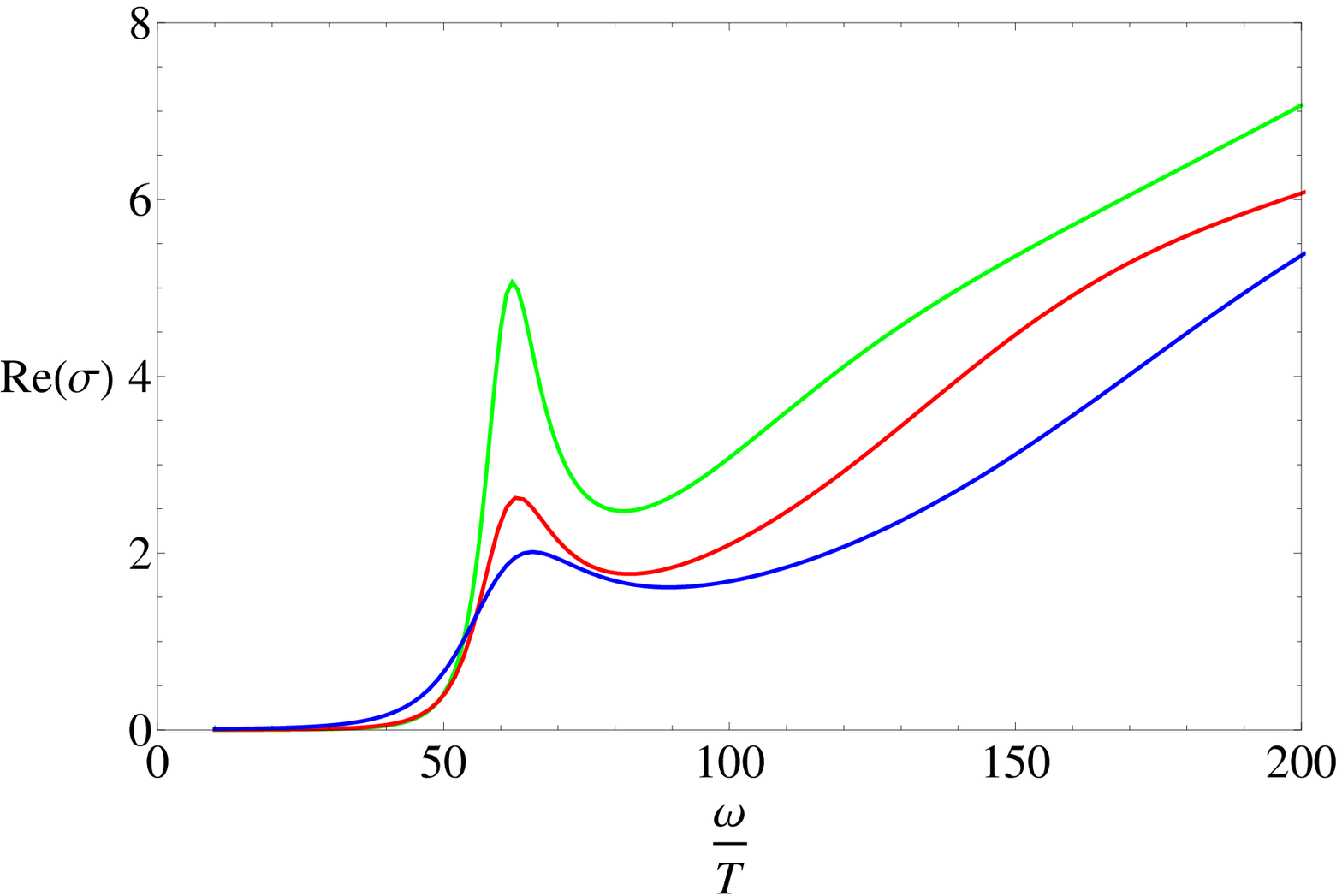}
\end{tabular}
  \caption{The real part of the conductivity as a function in terms of $\omega/T$ for the different values of $b$ and $\alpha$, at $d=5$, $T/T_c\approx0.47$ and  $m^2l^2_{\text{eff}}=-3$. The green, red and blue lines correspond to $b=0$, $0.05$, $0.1$. Left panel: $\alpha=0.02$. Right panel: $\alpha=0.06$.}\label{Re-conduc-d=5}
\end{figure}

\begin{figure}[t]
 \centering
\begin{tabular}{cc}
\includegraphics[width=0.45 \textwidth]{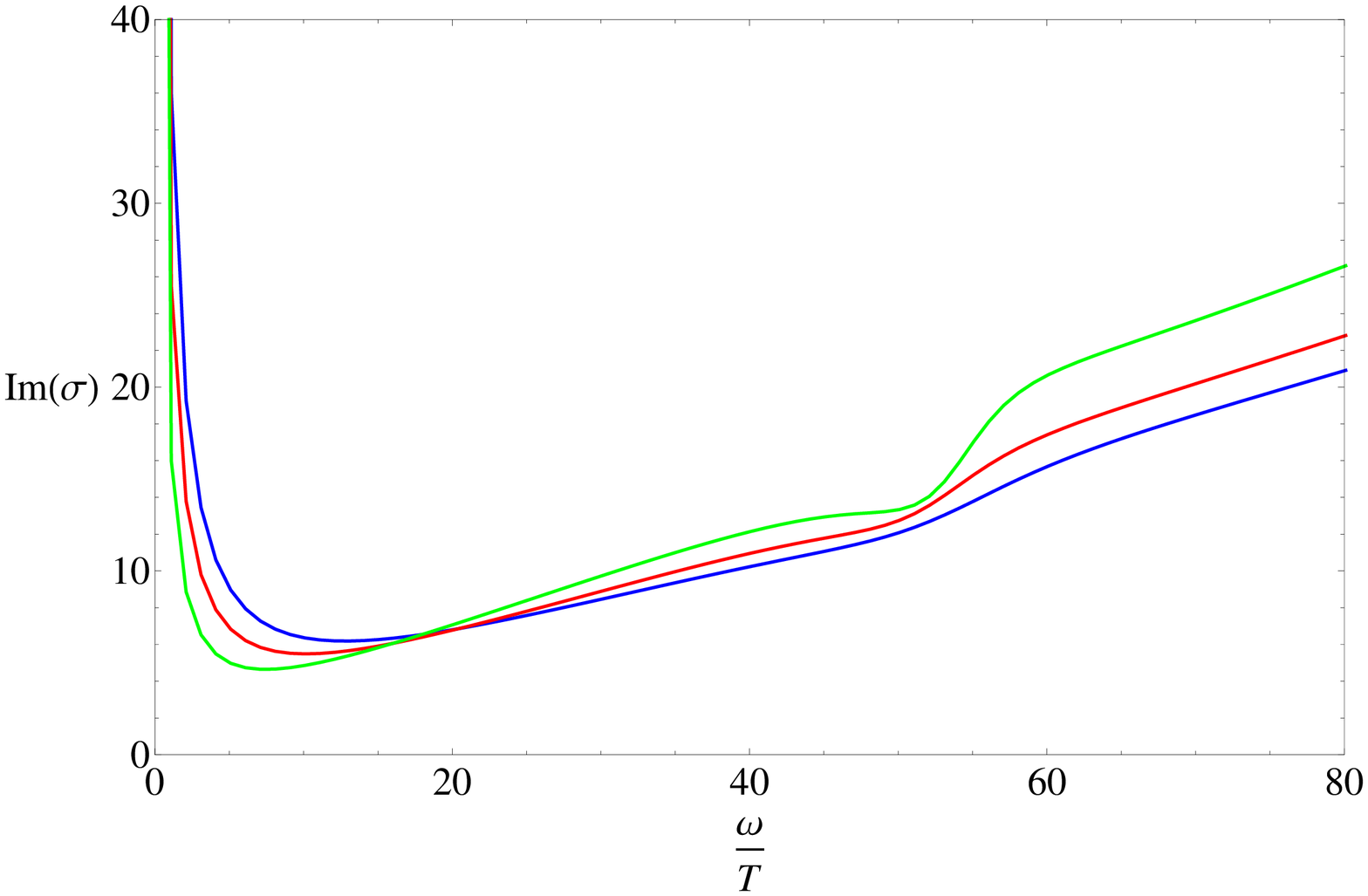}
\hspace*{0.05\textwidth}
\includegraphics[width=0.45 \textwidth]{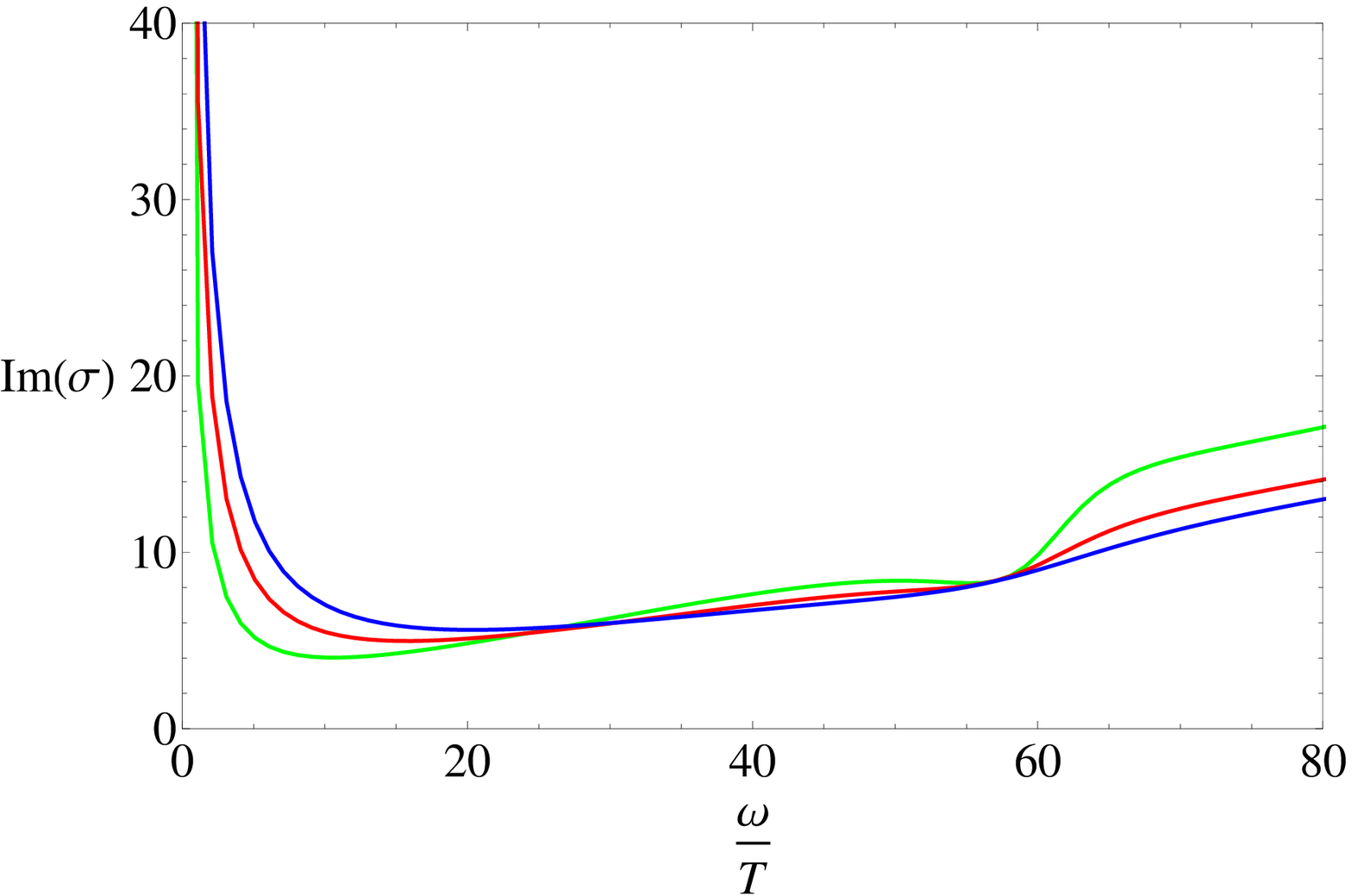}
\end{tabular}
  \caption{The behavior of the imaginary part of the conductivity in terms of $\omega/T$ for the different values of $b$ and $\alpha$, at $d=5$, $T/T_c\approx0.47$ and  $m^2l^2_{\text{eff}}=-3$. The green, red and blue lines correspond to $b=0$, $0.05$, $0.1$. Left panel: $\alpha=0.02$. Right panel: $\alpha=0.06$.}\label{Im-conduc-d=5}
\end{figure}
The numerical results for the behavior of the real and imaginary parts of the conductivity in terms of $\omega/T$, under the different values of $T$, $b$ and $\alpha$, are given in Figs. \ref{Re-conduc-d=5-alpha=0.06}, \ref{Im-conduc-d=5-alpha=0.06}, \ref{Re-conduc-d=5}, and \ref{Im-conduc-d=5}. As seen in Fig. \ref{Re-conduc-d=5-alpha=0.06}, below the critical temperature, there has the formation of a frequency gap in the real part of the conductivity. This points to the formation of a superconducting energy gap in the spectrum of charged excitations. The superconducting energy gap should be larger when decreasing the temperature. This is because the superconducting energy gap is relevant to the minimum energy which needs to break the condensation that becomes stronger as the temperature or the thermal motion (which prevents the formation of the bound state) decreases. Fig. \ref{Re-conduc-d=5} shows the effects of the nonlinear electrodynamics and the GB term on the superconducting energy gap. We can see that as increasing the nonlinear parameter, the superconducting energy gap becomes smaller, which is similar to the effect of the Born-Infeld electrodynamics. Whereas, the superconducting energy gap enlarges when the GB parameter increases. Also, for the region of the large frequency, the behavior of the real part of the conductivity obeys the power law \citep{Tong2013}, $\text{Re}[\sigma]\varpropto\omega^{d-4}$, which shows a normal state. Figs. \ref{Im-conduc-d=5-alpha=0.06} and \ref{Im-conduc-d=5} show the pole at $\omega=0$ in the imaginary part of the conductivity. By the Kramers-Kronig relation, this indicates a delta function at $\omega=0$ in the real part of the conductivity.

\section{\label{conclu} Conclusion}
It is interesting that the low-energy limits of the string theory lead to not only the higher-order curvature corrections for Einstein gravity but also the higher-order derivative corrections for the Maxwell or linear electrodynamics which implies the nonlinear electrodynamics. Motivated by this, we have investigated $s$-wave holographic superconductors, in the probe limit, in the framework of Einstein-Gauss-Bonnet gravity and the exponential nonlinear electrodynamics. This nonlinear electrodynamics was introduced in Ref. \cite{Hendi2012} with the aim of obtaining the new charged BTZ black hole solutions. Using the Sturm-Liouville eigenvalue method, we have analytically obtained the expression of the critical temperature in terms of the charge density and the spacetime dimension. The presence of the nonlinear parameter in the expression of the critical temperature is showed indirectly. It is observed that the larger nonlinear or GB parameter leads to the lower critical temperature. Whereas, the effect of the spacetime dimension on the critical temperature depends on the region of the charge density. More specifically, the superconducting phase transition in the higher (lower) dimensional systems can occur at the higher critical temperature at the efficiently low (high) charge density. It is showed that the analytic results are in very good agreement with the numerical results using the shooting method. 

Also, we have analytically calculated the condensation value and the critical exponent which is $1/2$ independently on the nonlinear electrodynamics and GB term. It is found that the condensation value gets larger as increasing the nonlinear parameter, the GB parameter, and the spacetime dimension. This means that the nonlinear electrodynamics, the GB correction, and the higher sapcetime dimension make the condensation harder. Finally, we have studied the optical conductivity whose real part indicates, below the critical temperature, a superconducting energy gap in the spectrum of charged excitations. It is observed that the superconducting energy gap becomes larger when increasing GB parameter. But, as increasing the nonlinear parameter it becomes smaller.

We would like to note that, in this paper, we only investigate $s$-wave holographic superconductors in the probe limit. However, it is interesting to study the effects of the backreaction of the scalar and gauge fields on the black hole spacetime background. In addition, based on the symmetry of the spatial part of the wave function of the bound state, superconductors can be classified as the $s$-wave, $p$-wave superconductors, etc. Since it is natural to extend this work to study $p$-wave holographic superconductors. These issues will be investigated in our future works.

\section*{Acknowledgments}
We would like to express sincere gratitude to the referees for their constructive comments and suggestions which have helped us to improve the quality of the paper.

\end{document}